\documentclass[a4paper,fleqn,usenatbib]{mnras}

\usepackage[T1]{fontenc}
\usepackage{ae,aecompl}

\usepackage{graphicx}	
\usepackage{amsmath}	
\usepackage{amssymb}	
\usepackage{ulem}       
\usepackage{cancel}     

\title[Transitioning MAD Disks]{GRRMHD Simulations of MAD Accretion Disks Declining from Super-Eddington to Sub-Eddington Accretion Rates}

\author[Brandon Curd \& Ramesh Narayan]{
Brandon Curd,$^{1,2}$\thanks{E-mail: brandon.curd@cfa.harvard.edu}
Ramesh Narayan$^{1,2}$
\\
$^{1}$ Harvard-Smithsonian Center for Astrophysics, 60 Garden Street, Cambridge, MA 02138, USA
\\
$^{2}$ Black Hole Initiative at Harvard University, 20 Garden Street, Cambridge, MA 02138, USA
\\
\\
}

\date{Accepted XXX. Received YYY; in original form ZZZ}

\pubyear{2022}

\begin{document}
\label{firstpage}
\pagerange{\pageref{firstpage}--\pageref{lastpage}}
\maketitle

\begin{abstract}
We present two general relativistic radiation magnetohydrodynamics (GRRMHD) simulations of magnetically arrested disks (MADs) around non-spinning ($a_*=0$) and spinning ($a_*=0.9$) supermassive black holes (BHs). In each simulation, the mass accretion rate is decreased with time such that we sample Eddington-scaled rates over the range $3 \gtrsim \dot{M}/\dot{M}_{\rm{Edd}}\gtrsim 0.3$. For the non-spinning BH model, the total and radiative efficiencies increase as the accretion rate decreases, varying over the range $\eta_{\rm{tot}}\sim9-16\%$ and $\eta_{\rm{rad}}\sim6-12\%$, respectively. This model shows very little jet activity. In contrast, the spinning BH model has a strong relativistic jet powered by spin energy extracted from the BH. The jet power declines with accretion rate such that $\eta_{\rm{jet}}\sim 18-39\%$ while the total and radiative efficiencies are $\eta_{\rm{tot}}\sim 64-100\%$ and $\eta_{\rm{rad}}\sim 45-79\%$, respectively. We confirm that mildly sub-Eddington disks can extract substantial power from a spinning BH, provided they are in the MAD state. The jet profile out to $100\, GM/c^2$ is roughly parabolic with a power-law index of $k\approx0.43-0.53$ during the sub-Eddington evolution. Both models show significant variability in the outgoing radiation which is likely associated with episodes of magnetic flux eruptions. The $a_*=0.9$ model shows semi-regular variations with a period of $\sim2000\, GM/c^3$ over the final $\sim10,000\, GM/c^3$ of the simulation, which suggests that magnetic flux eruptions may be an important source of quasi-periodic variability. For the simulated accretion rates, the $a_*=0$ model is spinning up while the $a_*=0.9$ model is spinning down. Spinup-spindown equilibrium of the BH will likely be achieved at $0.5 < a_{*,{\rm{eq}}} < 0.6$, assuming continuous accretion in the MAD state.

\end{abstract}




\section{Introduction}

Accretion disks around black holes (BHs) provide powerful engines for converting gravitational energy and BH spin energy into radiation emitted from the viscously heated inner accretion flow or from a powerful jet. In essence, BHs are engines that convert the accreting mass into other forms of energy (radiation, kinetic energy, magnetic power). The efficiency of this energy conversion depends intimately on the accretion disk properties, with one of the key components determining the energetics being the mass accretion rate itself (e.g see \citealt{2013LRR....16....1A,2014ARA&A..52..529Y}). Various models of accretion disks on the low and high end of accretion rates have found success in applications to various astrophysical phenomena involving BHs. When disks are accreting at super-Eddington rates, models such as geometrically thick, radiatively inefficient slim disks may apply \citep{1988ApJ...332..646A}. For accretion rates in the range $\sim10^{-2}-0.5$ times the Eddington rate, a geometrically thin, radiatively efficient disk, formally known as a Novikov-Thorne (NT hereafter) thin disk \citep{1973blho.conf..343N,1973A&A....24..337S} is likely present. Meanwhile, when the accretion rate drops yet further, the disk evaporates and forms a two-temperature advection dominated accretion flow, which is again radiatively inefficient and geometrically thick \citep{1995ApJ...452..710N,1997ApJ...489..865E,1976ApJ...204..187S,1977ApJ...214..840I,1982Natur.295...17R}. 

How exactly accretion disks evolve from one stage to another and how similarly the transitions operate across different BH mass scales and astrophysical settings is not well understood. Systems where the same BH experiences large variations in the mass accretion rate $\dot{M}$ as a function of time are particularly valuable in this regard. Accretion flows corresponding to stellar-mass BH X-ray binaries (BHXBs) and tidal disruption events (TDEs) around supermassive BHs (SMBHs), for example, show large changes in $\dot{M}$, and likely undergo state transitions. In both cases, for some fraction of the time the accretion rate is modestly sub-Eddington and the accretion disk is likely in the NT thin disk state.

Many AGN launch powerful astrophysical jets. One of the most attractive models for powering these jets is the Blandford-Znajek (BZ hereafter, \citealt{1977MNRAS.179..433B,1969NCimR...1..252P}), which requires that the BH has a relatively high spin and the BH is threaded by a large scale poloidal field. However, how exactly BHs in AGN both acquire and maintain this magnetic flux is not clear. In particular, poloidal fields in thin accretion disks are expected to rapidly diffuse away radially, indicating that thin accretion disks may not power BZ jets \citep{1994MNRAS.267..235L,1994ApJ...437..136L,1997ApJ...489..865E,1989ASSL..156...99V}. Additional support for this theoretical claim is provided by the jetted TDEs \textit{Swift} J1644+57 and \textit{Swift} J2058+05, whose X-ray flux showed a drastic decrease of more than two orders of magnitude several months after the initial detection, falling below detection limits \citep{2013ApJ...767..152Z,2015ApJ...805...68P}. Assuming both TDE events are the result of a SMBH disrupting a main sequence star, this rapid decline is thought to have occurred due to some critical accretion rate having been reached where the jet shuts off. At the time of drastic X-ray decline, both systems were likely sub-Eddington, which suggests that the disk may have indeed become geometrically thin \citep{2014MNRAS.437.2744T,2015ApJ...805...68P}.

How do jetted TDEs (and similar BH accretion systems) transition to a quiescent state where no apparent jet is observed? In terms of geometrically thick accretion disks, the magnetic flux and BH spin provide an explanation. In simple terms, accretion disks threaded by weak magnetic fields accrete via turbulence fueled by the magneto-rotational instability (MRI, \citealt{1991ApJ...376..214B}) and are expected to produce only mildly relativistic outflows at best \citep{2015MNRAS.453.3213S,2019MNRAS.483..565C}. Meanwhile, accretion disks threaded by powerful poloidal fields are capable of launching powerful jets \citep{2019MNRAS.483..565C}, presumably through the BZ process. It is common to refer to disks with weak fields as ``Standard and Normal Evolution'' (or SANE, \citealt{2012MNRAS.426.3241N}) disks and those with powerful poloidal fields as ``Magnetically Arrested Disks'' (or MAD, see \citealt{2003PASJ...55L..69N,2003ApJ...592.1042I}).

Although MADs provide an avenue for jet launching, theoretical studies, as discussed earlier, suggests that geometrically thin accretion disks cannot sustain a powerful jet. This fact led \citet{2014MNRAS.437.2744T} to conclude that the disk in \textit{Swift} J1644+57 must have become geometrically thin and simultaneously shed its magnetic flux and become non-MAD leading up to the X-ray shutoff. However, the assumption that the magnetic field diffuses away from the BH horizon once the disk becomes geometrically thin requires some scrutiny. Several works which simulated sub-Eddington accretion disks \citep{2016MNRAS.459.4397S,2016MNRAS.462..636A,2018MNRAS.480.3547M,2022arXiv220103526L} find that disks with $H/R \sim 0.1$ can be stabilized against thermal collapse via powerful magnetic fields. Simulations of sub-Eddington MADs \citep{2016MNRAS.462..636A,2018MNRAS.480.3547M,2022arXiv220103526L} did not show significant loss of magnetic flux over the simulated time. If the disk in \textit{Swift} J1644+57 was indeed MAD, and was stabilized against collapse when it went sub-Eddington by strong magnetic fields, then why did the X-rays decline so suddenly, signifying that the jet shut off?

Models of AGN often assume a thin disk and the presence of some corona, which produces significant X-ray emission through both direct emission and photons which scatter off of the accretion disk. One possible source of the corona is the jet/wind launched by the accretion flow. In particular, for spinning BHs, the BZ process extracts power from the BH \citep{2010ApJ...711...50T} at a rate:
\begin{equation}
    P_{\rm{BZ}} = \dfrac{\kappa}{16\pi c}\omega_H^2 \Phi_{\rm{BH}}^2f(\omega_H). \label{eq:PBZ}
\end{equation}
Here $\Phi_{\rm{BH}}$ is the magnetic flux threading the BH horizon, $\kappa$ is a constant related to the field geometry and is $0.044$ for a parabolic field structure, $\omega_H = a_*/[1 + (1-a_*^2)^{1/2}]$ is the dimensionless rotational frequency of the BH horizon, and $f(\omega_H) \approx 1$ for the spins $a_*=0, ~0.9$ that we consider in this work. Normalizing by the mass accretion rate $\dot{M}$, the efficiency is:
\begin{equation} \label{equ:etaBZ}
    \eta_{\rm{BZ}} \equiv \dfrac{\langle P_{\rm{BZ}} \rangle}{\langle\dot{M}\rangle c^2} = \dfrac{\kappa}{16\pi c} \omega_H^2 \langle\phi_{\rm{BH}}^2\rangle f(\omega_H),
\end{equation}
where the normalized flux $\phi_{\rm BH}$ is defined in equation~\eqref{eq:eq20}.
Thus one expects a jet to be present so long as the BH has some rotation and the disk has sufficient magnetic flux $\Phi_{\rm{BH}}$. This jet will likely source X-ray emission.

Scaling relations from simulations suggests that thin disks maintain lower magnetic flux. Geometrically thick disks reach a magnetically arrested state once the normalized magnetic flux at the BH horizon (see Section \ref{sec:definitions} for definitions) is $\phi_{\rm{BH}}\approx 70(1 - 0.38\omega_H)[(H/R)/0.3]$ \citep{2011MNRAS.418L..79T,2012JPhCS.372a2040T}. Here  $H/R$ is the disk scale height. If a similar scaling applies as the accretion rate $\dot{M}$ and the disk scale height $H/R$ decline, even sub-Eddington MAD disks may maintain substantial magnetic flux provided $H/R$ does not become too small. The powerful magnetic field in this scenario provides pressure support and also leads to a jet (and thus X-ray emission) for spinning BHs. With these considerations in mind, the rapid decline in the X-ray of \textit{Swift} 1644+57 is quite puzzling when interpreted as the result of a MAD accretion disk around a SMBH.

To further explore the viability of the MAD state in geometrically thin disks and to understand its relation to relativistic jets, in this work we study the properties of moderately thin, MAD accretion disks where the mass accretion rate is steadily decreased from super-Eddington rates to sub-Eddington rates. The systems described in this work are moderately thin ($H/R\approx0.15$), MAD accretion disks around a SMBH of mass $10^6\, M_\odot$ where the accretion rate peaks at $3-6 \, \dot{M}_{\rm{Edd}}$ and decreases to $\approx0.3 \, \dot{M}_{\rm{Edd}}$ by the end of the simulation. We perform one simulation with BH spin $a_*=0$ and one with $a_*=0.9$  to study the effects of BH spin. We study the efficiency, variability, jet structure, and emission as the disk evolves.

We describe our numerical methods and choice of initial conditions in Section \ref{sec:nummethods}. We also describe various diagnostics used to study each model. In Section \ref{sec:results}, we describe the time evolution of each model. We then analyze the outflow properties and measure the jet profile. We also perform a cross-correlation analysis to quantify the source of variability in the escaping radiation. In Section \ref{sec:spindown}, we quantify the spin evolution of each model as a function of accretion rate. In Section \ref{sec:discussion}, we compare our results with related work in the literature, and also consider observational implications. We summarize our findings in Section \ref{sec:conclusions}.

\section{Numerical Methods} \label{sec:nummethods}

The simulations presented in this work were performed using the general relativistic radiation magnetohydrodynamical (GRRMHD) code \verb=KORAL= \citep{2013MNRAS.429.3533S,2014MNRAS.439..503S,2017MNRAS.466..705S,2014MNRAS.441.3177M} which solves the conservation equations in a fixed, arbitrary spacetime using finite-difference methods. We solve the following conservation equations:
\begin{align}
  (\rho u^\mu)_{;\mu} &= 0, \label{eq:eq5} \\
  (T^\mu_\nu)_{;\mu} &= G_\nu, \label{eq:eq6} \\
  (R^\mu_\nu)_{;\mu} &= -G_\nu, \label{eq:eq7} 
\end{align}
where $\rho$ is the gas density in the comoving fluid frame, $u^\mu$ are the components of the gas four-velocity as measured in the ``lab frame'', $T^\mu_\nu$ is the MHD stress-energy tensor in the ``lab frame'',
\begin{equation} \label{eq:eq9}
  T^\mu_\nu = (\rho + u_g+ p_g + b^2)u^\mu u_\nu + \left(p_g + \dfrac{1}{2}b^2\right)\delta^\mu_\nu - b^\mu b_\nu,
\end{equation}
$R^\mu_\nu$ is the stress-energy tensor of radiation, and $G_\nu$ is the radiative four-force which describes the interaction between gas and radiation \citep{2013MNRAS.429.3533S,2014MNRAS.439..503S}. Here $u_g$ and $p_g=(\gamma - 1)u_g$ are the internal energy and pressure of the gas in the comoving frame and $b^\mu$ is the magnetic field four-vector which is evolved following the ideal MHD induction equation \citep{2003ApJ...589..444G}. 

The radiative stress-energy tensor is obtained from the evolved radiative primitives, i.e. the radiative rest-frame energy density and the four velocity of this frame, using the M1 closure scheme modified by the addition of radiative viscosity \citep{2013MNRAS.429.3533S,2015MNRAS.447...49S}.

The interaction between gas and radiation is described by the radiation four-force $G_\nu$. The opposite signs of this quantity in the conservation equations for gas and radiation stress-energy (equations \ref{eq:eq6}, \ref{eq:eq7}) reflect the fact that the gas-radiation interaction is conservative, i.e. energy and momentum are transferred between gas and radiation. For a detailed description of the four-force see \citet{2017MNRAS.466..705S}. We include the effects of scattering via the electron scattering opacity ($\kappa_{\rm{es}}$). We also account for absorption and emission via the free-free absorption opacity ($\kappa_{\rm{a}}$) and bound-free absorption opacity through the Sutherland Dopita model \citep{1993ApJS...88..253S}. We assume a Solar metal abundance for the gas such that the Hydrogen mass-fraction $X=X_\odot=0.7381$, the Helium mass-fraction $Y=Y_\odot=0.2485$, and the mass-fraction of all other elements $Z=Z_\odot=0.0134$. We also include the effects of thermal synchrotron and Comptonization \citep{2015MNRAS.454.2372S,2017MNRAS.466..705S}.

We quantify the resolution of the fastest growing mode of the MRI by computing the quantities:
\begin{align}
  Q_\theta = \dfrac{2\pi}{\Omega dx^\theta}\dfrac{|b^\theta|}{\sqrt{4\pi\rho}}, \label{eq:eq10} \\
  Q_\phi = \dfrac{2\pi}{\Omega dx^\phi}\dfrac{|b^\phi|}{\sqrt{4\pi\rho}}, \label{eq:eq11}
\end{align}
where $dx^i$ (the size of the grid cell) and $b^i$ (the magnetic field) are both evaluated in the orthonormal frame, $\Omega$ is the angular velocity, and $\rho$ is the gas density. Numerical studies of the MRI have shown that values of $Q_\theta$ and $Q_\phi$ in excess of at least 10 are needed to resolve the fastest growing mode \citep{2011ApJ...738...84H}.

We use modified Kerr-Schild coordinates with the metric corresponding to a BH with mass $10^6\, M_\odot$ and two values of spin, $a_*=0,\,0.9$. In each simulation, we run the initial inflow of gas in 2D ($r,\theta$) coordinates with a resolution $N_r \times N_\theta=256\times128$. We then pause the simulation and perform a regrid (which we later refer to as `regrid 1') to 3D ($r,\theta,\phi$) with a resolution of $N_r \times N_\theta \times N_\phi=256\times128\times128$ and evolve the simulation until the accretion rate approaches sub-Eddington levels. At this time, we again pause the simulation and perform an additional regrid (which we later refer to as `regrid 2') in which we double the resolution in $\theta$ with a resolution of $N_r \times N_\theta \times N_\phi=256\times256\times128$ and evolve the simulation until the accretion rate approaches $~\sim0.3$ times the Eddington limit. We set the inner edge of the domain such that the inner six domain cells are inside the BH horizon, which guarantees that the interior grid boundary is causally disconnected from the exterior. The radial grid cells are spaced logarithmically in radius and the cells in polar angle $\theta$ have smaller sizes $\Delta \theta$ towards the equatorial plane (see \citealt{2011MNRAS.418L..79T}). The cells are equally spaced in azimuth. At the inner radial boundary ($R_{\rm{min}}$), we use an outflow condition while at the outer boundary ($R_{\rm{max}}$) we use a similar boundary condition and in addition prevent the inflow of gas and radiation. At the polar boundaries, we use a reflective boundary. The polar boundaries are radius dependent such that near the horizon the minimum/maximum polar coordinate is shifted slightly away from the pole to reduce the minimum time step, while at larger radii the grid extends closer to the poles \citep{2011MNRAS.418L..79T}. We employ a periodic boundary condition in azimuth. For full 3D portions of each simulation, the grid covers $-\pi \leq \phi \leq \pi$. 

\begin{table*}
    \centering
    \begin{tabular}{c c c c c c c }
        \hline
        \hline 
        Model Name & $a_*$ & $R_{\rm{in}}$ & $R_{\rm{max}}$ & $R_{\rm{inj}}$ & Resolution & Duration \\
                & & $(r_g)$ & $(r_g)$ & $(r_g)$ & $N_r \times N_\theta \times N_\phi$ & ($t_g$)\\
        \hline
        \texttt{KIa00} & 0 & $1.5$ & $10^5$ & $40$ &  $256\times256\times128$  & 29,000 \\
        \texttt{KIa09} & 0.9 & $1.1$ & $10^5$ & $40$ & $256\times256\times128$ & 34,000\\
    \hline
    \end{tabular}
    \caption{Description of simulations presented in this work. Note that we give the final resolution of each simulation. Both simulations were performed in 2D, and 3D with multiple resolution increases. We describe this in detail in Section \ref{sec:diskinjection}.}
    \label{tab:tab1}
\end{table*}

We performed two simulations, one with zero spin $a_*=0$ (\texttt{KIa00}) and one with near maximal spin $a_*=0.9$ (\texttt{KIa09}). In this naming convention \texttt{`KI'} identifies these models as Keplerian disk injection models (as described in Section \ref{sec:diskinjection}, the gas in these simulations is injected at a radius $R_{\rm{inj}}$) and \texttt{`a00'} or \texttt{`a09'} identifies the BH spin. Details of each simulation are given in Table \ref{tab:tab1}.

\subsection{Definitions} \label{sec:definitions}

Throughout this work, we use gravitational units to describe physical parameters. For distance we use the gravitational radius $r_g\equiv GM_{\rm{BH}}/c^2$ and for time we use the gravitational time $t_g\equiv GM_{\rm{BH}}/c^3$, where $M_{\rm{BH}}$ is the mass of the BH. Often, we set $G=c=1$, so the above relations would be equivalent to $r_g=t_g=M_{\rm BH}$\footnote{For a BH mass of $10^6\,M_\odot$, the gravitational radius and time in cgs units are $r_g = 1.48\times 10^{11}$ cm and $t_g = 4.94$ s, respectively.}. Occasionally, we restore $G$ and $c$ when we feel it helps to keep track of physical units.

We adopt the following definition for the Eddington mass accretion rate:
\begin{equation} \label{eq:mdotEdd}
  \dot{M}_{\rm{Edd}} = \dfrac{L_{\rm{Edd}}}{\eta_{\rm NT} c^2},
\end{equation}
where $L_{\rm{Edd}} = 1.25\times 10^{38}\, (M_{\rm{BH}}/M_\odot)\, {\rm erg\,s^{-1}}$ is the Eddington luminosity, $\eta_{\rm{NT}}$ is the radiative efficiency of a NT thin disk around a BH with spin parameter $a_*$,
\begin{equation} \label{eq:etaNT}
  \eta_{\rm{NT}} = 1 - \sqrt{1 - \dfrac{2}{3 r_{\rm{ISCO}}}}.
\end{equation}
In the above expression, $r_{\rm{ISCO}}=3+Z_2 - \sqrt{(3-Z_1)(3+Z_1+2Z_2)}$ is the radius of the Innermost Stable Circular Orbit (ISCO, \citealt{1973blho.conf..343N}) in the Kerr metric, where $Z_1 = 1 + (1-a_*^2)^{1/3}\left((1+a_*)^{1/3}+(1-a_*)^{1/3}\right)$ and $Z_2 = \sqrt{3a_*^2 + Z_1^2}$. For $a_* =0.0$, the efficiency is $\eta_{\rm{NT}}=5.70\%$ while for $a_* =0.9$ the efficiency is $\eta_{\rm{NT}}=15.58\%$.

In this section, we provide brief descriptions of quantities used to study the simulation. We compute the net mass inflow rate as:
\begin{equation} \label{eq:mdotin}
  \dot{M}(r) = -\int_0^\pi \int_0^{2\pi} \sqrt{-g}\,\rho \,u^r d\phi d\theta.
\end{equation}
We treat the accretion of mass onto the BH as this integral taken at the horizon $r_H$, and for simplicity we define $\dot{M}\equiv \dot{M}(r_H)$ throughout the text. We also define the time averaged Eddington ratio (with time averages over $3000\,t_g$):
\begin{equation}
  \lambda \equiv \dfrac{\langle\dot{M}\rangle}{\dot{M}_{\rm{Edd}}},
\end{equation}
which we use to differentiate regions of super-Eddington ($\lambda > 1$) and sub-Eddington ($\lambda < 1$) accretion states. We commonly refer to $\lambda$ to describe stages in each simulation. We also consider the net mass outflow rate at radius $r$:
\begin{equation} \label{eq:mdotin}
  \dot{M}_{\rm{out}}(r) = \int_0^\pi \int_0^{2\pi} \sqrt{-g}\,\rho \,{\rm{max}}(u^r,0) d\phi d\theta.
\end{equation}

The magnetic flux is computed as:
\begin{equation} \label{eq:eq20}
  \Phi = \dfrac{1}{2} \int_0^{\pi}\int_{0}^{2\pi}|B^r(r)|\sqrt{-g}\, d\phi d\theta,
\end{equation}
where $B^r(r)$ is the radial component of the magnetic field in Gaussian units. We quantify the magnetic field strength at a given radius through the normalized magnetic flux parameter \citep{2011MNRAS.418L..79T}:
\begin{equation} \label{eq:eq20}
  \phi(r) = \dfrac{\Phi(r)}{\sqrt{\dot{M}(r)}}.
\end{equation} 
Throughout, we use $\phi_{\rm{BH}}\equiv \phi(r_H)$ to describe the normalized magnetic flux threading the BH. 

To characterize the properties of the gas, we use the total pressure $p_{\rm{tot}} \equiv p_{\rm{gas}} + p_{\rm{rad}} + p_{\rm{mag}}$, where the individual terms are the gas pressure ($p_{\rm{gas}}$), radiation pressure ($p_{\rm{rad}}$), and magnetic pressure ($p_{\rm{mag}}$). It is also useful to define the magnetic pressure ratio 
\begin{equation}
  \beta \equiv \dfrac{(p_{\rm{gas}} + p_{\rm{rad}})}{p_{\rm{mag}}}.
\end{equation}

We quantify the disk thickness by estimating the density scale height:
\begin{equation} \label{eq:eqHR}
    \dfrac{H}{R} = \sqrt{\dfrac{\int \int \rho \sqrt{-g} \, \, | \pi/2 - \theta |^2 \, d\theta d\phi}{\int \int \rho \sqrt{-g} \,d\theta d\phi}.}
\end{equation}
The above integral is performed over all angles $\theta$ and $\phi$.

We estimate the electron scattering photosphere location for an observer at infinity along the direction $(\theta,\phi)$ by integrating the optical depth radially inward from the outer boundary of the grid. Far from the BH, the curvature of spacetime is negligible, so we simply integrate at constant $(\theta,\phi)$ in the ``lab frame'':
\begin{equation} \label{eq:taues}
  \tau_{\rm{es}}(r) = -\int_{R_{\rm{max}}}^r \dfrac{\rho \kappa_{\rm{es}}}{c} \left(u^t - u^r\right)\sqrt{g_{rr}}\,dr',
\end{equation}
where $\kappa_{\rm{es}} = 0.2(1+X)\kappa_{\rm{KN}}\,{\rm cm^2}$ is the electron scattering opacity, $\kappa_{\rm{KN}}$  is the Klein-Nishina correction factor for thermal electrons \citep{2017MNRAS.466..705S}, and $R_{\rm{max}}$ is the radius corresponding to the outer boundary of the grid. For the gas and radiation temperatures in the simulations presented here, the Klein-Nishina correction is negligible and the electron scattering opacity is essentially equal to the Thomson opacity. In this work, we choose the location of the photosphere as the $\tau_{\rm{es}}=1$ surface.

We also characterize the accretion flow using the magnetization parameter, $\sigma_M\equiv b^2/\rho$, which quantifies the magnetic energy to rest mass energy ratio. We define the magnetized jet as the region where $\sigma_M > 1$. To maintain numerical stability, we introduce mass in the magnetized jet using a ceiling condition on the magnetization $\sigma_M \leq 100$ throughout each simulation. In measuring the jet profile, we make use of the $\sigma_M =1$ contour. However, when computing the total energy flux in the jet, we use a different definition of the jet boundary, based on the Bernoulli parameter. We prefer this definition since gas which may be accelerated to highly relativistic asymptotic speeds ($\gamma_{\infty} \gtrsim 2$) will lie in the $\sigma_M>1$ region. In addition, the Bernoulli parameter profiles are less well defined at large radii and are not useful for computing the jet shape. However, a significant fraction of gas which is too fast ($\gamma_{\infty} > 1.05$) to be called a `wind' lies outside of the $\sigma_M>1$ region so we make use of the Bernoulli definition to compute the total energy of the jet.

We are particularly interested in the energy outflows. To that end, we define the outgoing radiative bolometric luminosity by:
\begin{equation} \label{eq:eq15}
  L_{\rm{bol}}(r_{\rm{lum}}) = \int_{0}^{\pi}\int_0^{2\pi}\sqrt{-g}\, \,{\rm{max}}(-R^r_{\ \, t},0) d\phi d\theta,
\end{equation}
which gives the flux of outgoing radiation energy through a surface at radius $r_{\rm{lum}}=\max(30\, r_g, r_{\tau_{\rm{es}}=1})$. By this definition, the default $r_{\rm{lum}}=30\, r_g$, which lies inside of the injection radius $R_{\rm{inj}}$, but it moves out to the photosphere if the latter lies outside $30\, r_g$.

The MHD energy flux is computed as:
\begin{equation} \label{eq:ltot}
  L_{\rm{MHD}}(r) = -\int_{\theta_1}^{\theta_2}\int_0^{2\pi}\sqrt{-g}\, (T^r_{\ \, t} + \rho u^r) d\phi d\theta,
\end{equation}
where we integrate the radial flux of energy carried by gas plus magnetic field ($T^r_{\ \, t}$) and subtract out the rest-mass energy ($\rho u^r$, since it does not lead to observational consequences for an observer at infinity). Note that the integral is only over $\theta_1 \leq \theta \leq \theta_2$ since we consider the energy outflow in jet and wind seperately. We define these regions below. When measured at large radii, this quantity determines the total outflowing MHD energy extracted from the accretion disk.


We define the accretion flow as three distinct regions (disk, wind, and jet) based on the total energy via the Bernoulli parameter. We also make use of the electron scattering opacity, $\tau_{\rm{es}}$, to determine where the region of the fluid under consideration is optically thin or thick. In optically thick regions ($\tau_{\rm{es}} \geq 1$), the radiation is advected with the flow and can contribute to acceleration of the gas so we treat it as contributing to the Bernoulli parameter. Meanwhile, in optically thin regions ($\tau_{\rm{es}} < 1$) we assume that only the MHD components are relevant to the total gas energy. Building on the definition given in \citet{2019MNRAS.483..565C}, we define a Bernoulli parameter which smoothly connects between optically thin and optically thick regions via:
\begin{equation} \label{eq:Bernoulli}
    \rm{Be} = 
    \begin{cases}
      -\dfrac{(T^t_{\ \, t} + \rho u^t) + (1 - \tau_{\rm{es}}^{-1/2})R^t_{\ \, t}}{\rho u^t}, \ \, \tau_{\rm{es}} \geq 1\\
      -\dfrac{(T^t_{\ \, t} + \rho u^t)}{\rho u^t}, \ \, \tau_{\rm{es}} < 1\\
    \end{cases}
\end{equation}
We have adopted this definition to account for the fact that radiative acceleration of the gas will be weak in the optically thin region. We define the `disk' region of the simulation as bound gas with $\rm{Be} < 0$. Both the `wind' and `jet' are unbound and are generally radially out flowing. The wind is defined as any fluid with $0 < \rm{Be}\leq 0.05$. The jet is any fluid with $\rm{Be}> 0.05$. This choice of cutoff for wind vs. jet is based on the velocity at infinity ($v_\infty$), with the wind having $v_\infty \lesssim 0.3 c$ and the jet having $v_\infty \gtrsim 0.3 c$.

We define the radiation, jet, and wind efficiencies as:
\begin{equation}
    \eta_{\rm{rad}} = \dfrac{1}{\dot{M}} L_{\rm{bol}}(r_{\rm{lum}}),
\end{equation}
\begin{equation}
    \eta_{\rm{jet}} = \dfrac{1}{\dot{M}} L_{\rm{MHD}}(r=30\,r_g, {\rm{Be}}\geq0.05),
\end{equation}
and
\begin{equation}
    \eta_{\rm{wind}} = \dfrac{1}{\dot{M}} L_{\rm{MHD}}(r=30\,r_g, 0.05 > {\rm{Be}} > 0).
\end{equation}
In each case, positive values correspond to energy being extracted from the system. 
We define the total efficiency as: 
\begin{equation}
    \eta_{\rm{tot}}=\eta_{\rm{jet}}+\eta_{\rm{wind}}+\eta_{\rm{rad}}.
\end{equation}

We are also interested in the total angular momentum flux flowing into the system:
\begin{equation} \label{eq:jtot}
  \dot{J}(r) = -\int_{0}^{\pi}\int_0^{2\pi}\sqrt{-g}\, (T^r_{\ \, \phi} + R^r_{\ \, \phi}) d\phi d\theta.
\end{equation}
The specific angular momentum flux into the BH is then 
\begin{equation} \label{eq:jdot}
  j(r) = \dfrac{\dot{J}(r)}{\dot{M}}.
\end{equation}
In this work, we adopt $r=r_H$ to measure the specific angular momentum into or out of the BH. If $j(r_H)>0$, angular momentum is being gained by the BH, and if $j(r_H)<0$, angular momentum is being lost by the BH. We follow \citet{2005ApJ...620...59S} and define the BH's dimensionless spin-up parameter:
\begin{equation} \label{eq:spinup}
    s\equiv \dfrac{d(a/M)}{dt}\dfrac{M}{\dot{M}}= j - 2a_*(1-\eta_{\rm{tot}}),
\end{equation}
where positive values correspond to the BH spin increasing and negative values correspond to the BH spin decreasing with time. More positive/negative values imply more rapid changes in the BH spin.

\subsection{Disk Injection} \label{sec:diskinjection}

\begin{figure}
    \centering{}
	\includegraphics[width=\columnwidth]{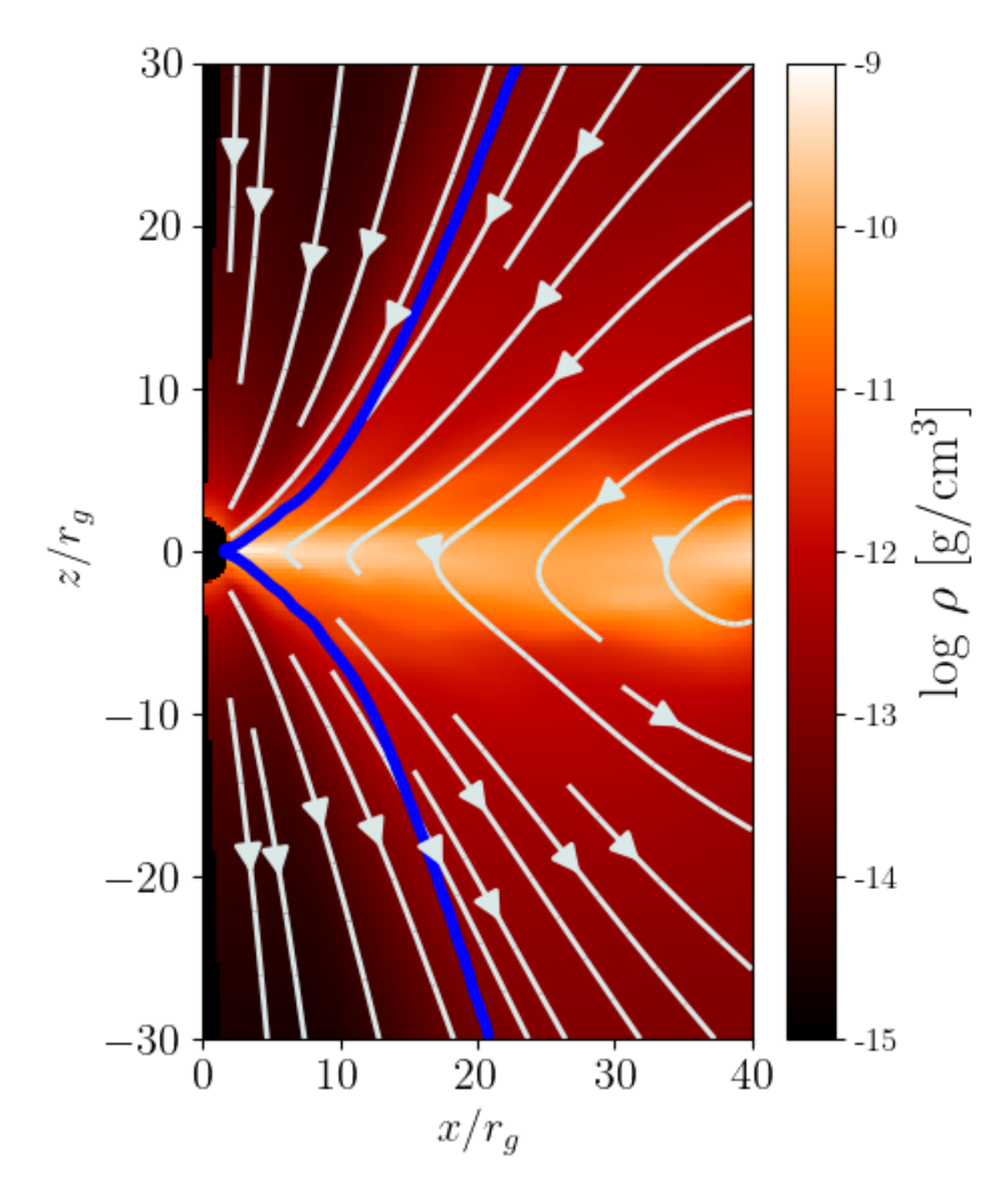}
    \caption{Example magnetic field at $t=15,000 \, t_g$ in model \texttt{KIa09}. The gas density of the disk (colors) is shown along with the magnetic field lines and the $\sigma_M=1$ contour (blue line), which delineates the highly magnetized region of the jet.}
    \label{fig:Bfield}
\end{figure}

The simulation domain is initialized with a low density atmosphere with a density profile that scales as $r^{-2}$ and has a maximum density of $6\times10^{-17}{\rm{\, g\, cm^{-3}}}$ at the horizon. The atmosphere is initialized with a constant radiation temperature of $T_{\rm{atm}}=10^5$ K. 

Using a similar scheme to \citet{2021MNRAS.507.3207C}, we introduce gas that is initially on a circular orbit into the simulation domain via an injection boundary condition in a small ring of cells at an injection radius $R_{\rm{inj}}=40\,r_g$ that is interior to the maximum radius ($10^5\, r_g$) of the simulation domain. The vertical extent of the injected disk is set to $(H/R)_0=0.05$, but as the gas first flows into the simulation domain (prior to the formation of a disk), it self-consistently puffs up due to the high magnetic and radiation pressure. The gas density in the injection cells varies with vertical distance from the mid plane ($\theta=\pi/2$) as $\rho \propto \left[1 - \left(\frac{|\theta-\pi/2|}{(H/R)_0}\right)^2\right]^3$. The normalization is set to give the desired mass injection rate, which we define later. The primary motivation for this approach is that it allows for precise control of the accretion rate as the simulation evolves in time so long as the timescale on which the accretion rate is varied is longer than the viscous accretion time from $R_{\rm{inj}}$ to the horizon $r_H$. The accretion time for a parcel of gas in a MAD, geometrically thick disk is approximately given by (in gravitational units):
\begin{equation}
    \dfrac{T_{\rm{acc}}}{t_g} \approx \dfrac{r}{0.1v_{\rm{Kep}}} = 10 r^{3/2},
\end{equation}
where $r$ is the radius, and $v_{\rm{Kep}} = r^{-1/2}$ is the Keplerian orbital velocity. Here we have assumed that the radial velocity of the inflowing gas is 10\% of the Keplerian velocity motivated by the findings of \citet{2012MNRAS.423.3083M}.

For this numerical experiment, we chose $R_{\rm{inj}}=40\, r_g$, which corresponds to an accretion time of $T_{\rm{acc}}\approx 2500\, t_g$.  We initialize the orbital properties of the gas by setting the radial velocity $v_r = 0$ and setting the angular momentum to that of a circular orbit at the injection radius.

A nominal gas temperature of $T_{\rm{inj}}=10^9$ K is set at the injection boundary. This is not the true temperature, but is a measure of the total pressure. We use the initial gas pressure obtained from $p'=p_{\rm{gas}}(T_{\rm{inj}})$ to split the internal energy into gas and radiation energy density by solving the condition $p' = p_{\rm{gas}}+p_{\rm{rad}}$ and finding a new gas and radiation temperature which assumes thermal equilibrium of the gas. This gives $T_{\rm{gas}}\approx5\times10^5$ K in the injection cells.

We set the magnetic field in the two radially interior injection boundary cells to an initially nearly vertical field geometry $B^\theta \propto \cos \left[\frac{\theta-\pi/2}{(H/R)_0}\right] $ and a magnetic pressure ratio of $\beta = 33$. This choice of $\beta$ leads to the rapid formation of a MAD disk as the gas begins to accrete.  We set $B^r = B^\phi = 0$ at the injection point. The divergence constraint $\nabla\cdot \bold{B}=0$ is enforced through the flux constrained transport algorithm introduced by \citet{2000JCoPh.161..605T}. We do not explicitly initialize the ghost cells such that $\nabla\cdot \bold{B}=0$ is guaranteed, but the magnetic field that ultimately gets injected with the gas has a divergence of the order $|\nabla\cdot \bold{B}|/|\bold{B}|\sim 10^{-7}$ throughout the evolution. We note that a similar magnetic field initialization was used in \citep{2022arXiv220111753K} to simulate a magnetized Bondi-Hoyle-Lyttleton flow; however, they inject gas from the outer boundary of the simulation. We show a snapshot of the resulting disk and field loops in Figure \ref{fig:Bfield}.

The injection boundary is four cells wide in the radial direction. At the two radially interior cells, we set the gas primitives using the quantities described above. Meanwhile, for the two radially exterior cells we employ a reflecting boundary condition in radius. This is to avoid any spurious inflows or outflows behind the injection point. Gas moving with $v^r<0$ at $r>40\, r_g,\, |\theta - \pi/2| \leq (H/R)_0$ encounters a reflective boundary and cannot flow further in.\footnote{We note that this does not affect the validity of the numerical solution of the inflow/wind/jet since the gas behind the injection point is outflowing after the disk forms.}

One difficulty with treating the injection of gas inside of the simulation domain is the choice of boundary conditions at the top and bottom ($\theta$) edges and the radial ($r$) edges. Enforcing reflection at both the exterior radius and the top/bottom edges in $\theta$ is non-trivial. For instance, setting the primitives in the ghost cells at the radial face to prevent inflow/outflow does not guarantee that no inflow/outflow will occur at the top/bottom angular corner cells. This is particularly pernicious with MAD disks as enforcing a reflecting boundary condition on the magnetic field can lead to unwanted inflow of gas carrying the opposite magnetic field polarity at the angular boundary. We sought to minimize mass from flowing out of the simulation domain through the injection region while maintaining the desired injected field topology at all times, so we opted to set the magnetic field $B^r=B^\theta=B^\phi = 0$ at the 2 radially exterior grid cells and enforce a reflecting boundary condition only in radius on the other gas primitives at the injection boundary. This guaranteed that no magnetic field is injected if mass is allowed to flow in through a corner cell. We do not believe this to be a major issue as the gas flowing around the boundary is radially outflowing throughout the simulation. In addition, we compute the simulation properties near the midplane at $r< R_{\rm{inj}}$ throughout this work where the effect of these boundary conditions should be minimal. We demonstrate in Appendix \ref{sec:appA} that either reflecting $\bold{B}$ or setting $\bold{B}=0$ at the 2 radially exterior cells does not impact the conclusions drawn from the simulation.

Here we further detail the regrid process and how the simulation evolves through each stage. As detailed in the previous subsection, we first inject the gas in 2D with a resolution of 256 cells in $r$ and 128 cells in $\theta$. We initially fix the injection rate to $\dot{M_0}=10 \dot{M}_{\rm{Edd}}$ until the magnetic flux at the horizon reaches $\phi_{\rm{BH}} > 40$. 

After this stage, which takes $10,000 \, t_g$ in each simulation, we perform `regrid 1' and copy the 2D data onto a fully 3D grid with 128 cells in polar coordinate $\phi$. We also introduce a 5\% perturbation in the azimuthal velocity before restarting the simulation to break the azimuthal symmetry. In \texttt{KIa00}, we continue injecting mass at a fixed rate until $t=15,000\, t_g$ and begin decreasing it after that time. For \texttt{KIa09}, we begin to decrease the mass injection rate at $t=10,000\, t_g$, immediately after the regrid. The mass inflow rate at the injection boundary decreases over time following a $\dot{M}\propto t^{-5/3}$ profile. The exact description of the mass injection is given by:
\begin{equation} \label{eq:mdotinj}
    \dot{M}_{\rm{inj}}(t) = \dot{M}_0 \left[\dfrac{(t-t_0)}{T_{\rm{acc}}(R_{\rm{inj}})} + 1\right]^{-5/3}
\end{equation}
where $\dot{M}_0$ is the peak mass inflow rate, $t_0$ is the time at which the decay starts, and $t$ is the time since the beginning of injection in gravitational units. The power-law dependence that we have adopted in Equation \ref{eq:mdotinj} is arbitrary and is merely adopted for its similarity to the time evolution in TDEs, whose mass fallback rate varies approximately as $\dot{M}_{\rm{fb}}\propto t^{-5/3}$. A similar result could be obtained by using a larger $T_{\rm{acc}}$ coupled with a shallower power-law as long as the disk has time to adjust to changes at the injection boundary. It is worth noting that $T_{\rm{acc}}\approx 2500\, t_g$ is substantially shorter than the decay times of $\sim 28,000-180,000\, t_g$ adopted in \citet{2021MNRAS.507.3207C}. In this work, our somewhat small choice of $T_{\rm{acc}}$ is motivated by the computational requirements to probe a range of accretion states that differ by roughly an order of magnitude over the duration of a single simulation.

After the disk becomes sub-Eddington, which happens around $21,000\, t_g$ for \texttt{KIa00} and $20,000\, t_g$ for \texttt{KIa09}, we perform `regrid 2' and double the resolution in $\theta$ from $N_\theta = 128$ to $N_\theta=256$ to better resolve the disk. In terms of resolution, the grid has $\Delta \theta \approx 0.005 \,(0.01)$ near the mid-plane after (before) the regridding. As a result, we resolve the disk with roughly 30 (15) cells per disk scale height (we provide the precise disk scale height later in the text), even at the lowest accretion rates we consider.

\begin{figure*}
    \centering{}
	\includegraphics[width=\textwidth]{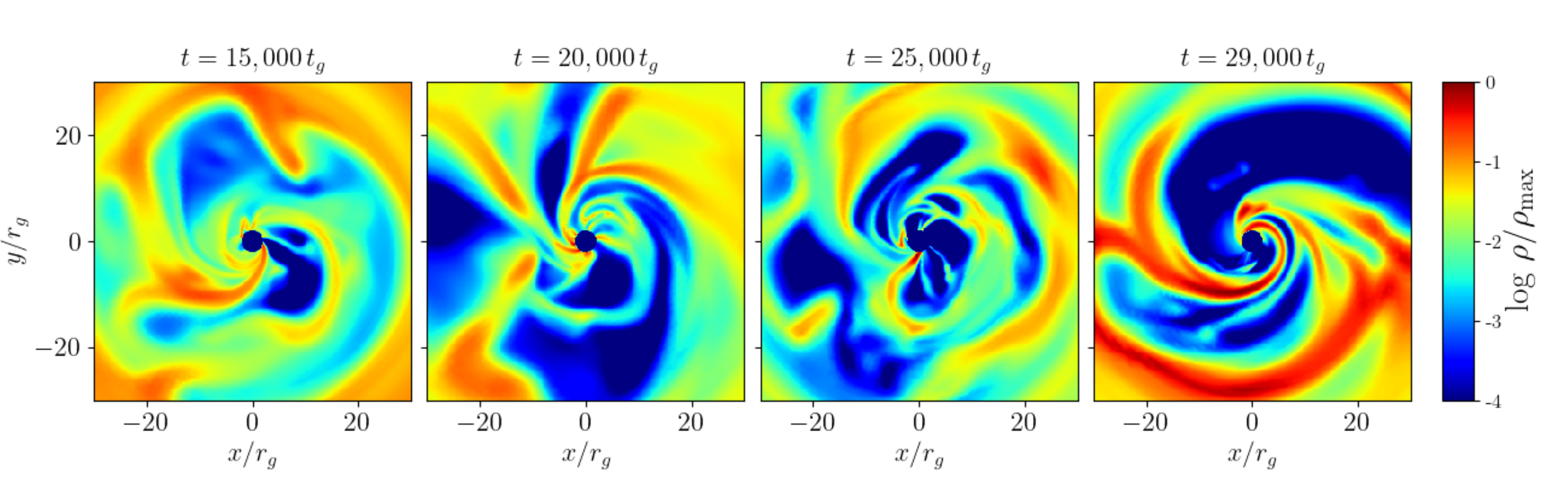}\\
	\includegraphics[width=\textwidth]{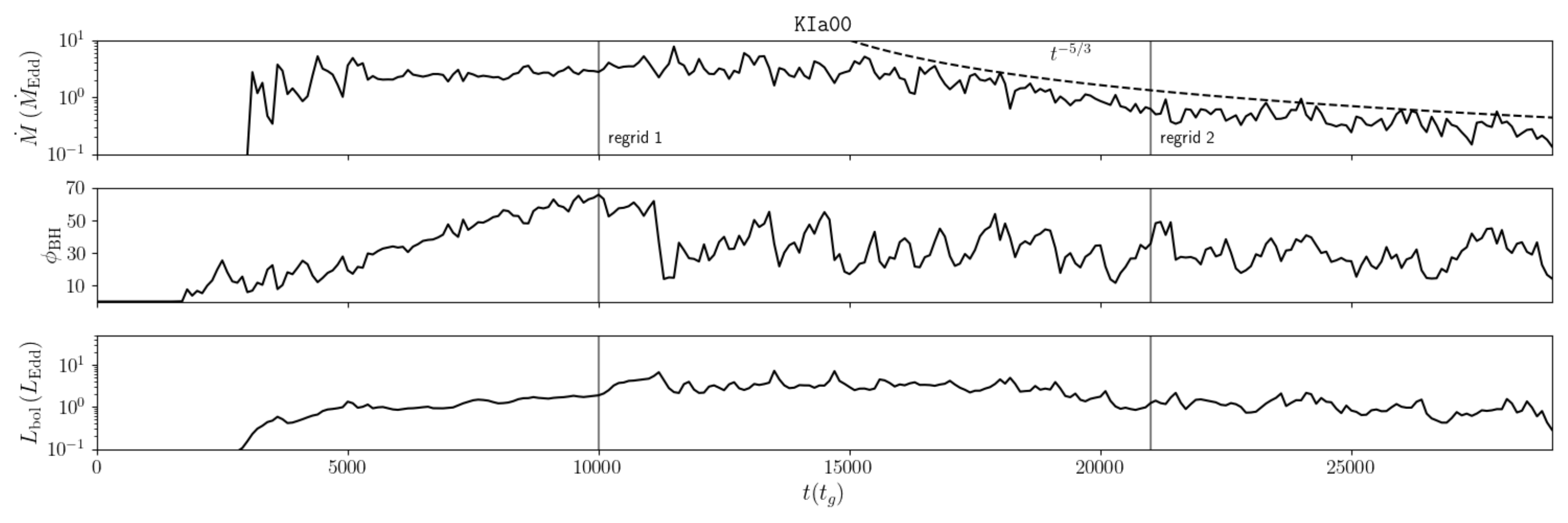}
    \caption{In the top row we show snapshots of the accretion disk in the midplane spanning the evolution of \texttt{KIa00} from $t=15,000-29,000\, t_g$ in a $60r_g\times 60 r_g$ region centered on the black hole. The gas density, scaled by the maximum density in the disk ($\rho/\rho_{\rm{max}}$), is indicated by colors in each panel. Gas primarily flows in along several spiral arms and the low density regions (blue) of highly magnetized gas, sometimes referred to as `magnetic islands,' can sometimes make up a significant fraction of the surface area in the disk. The full simulation history is shown in the bottom panels. We show the time evolution of accretion rate $\dot{M}$, normalized magnetic flux at the BH horizon $\phi_{\rm{BH}}$, and bolometric luminosity $L_{\rm{bol}}$. The vertical gray lines indicate the time at which we perform the regrid from 2D to 3D (regrid 1) and when we double the $\theta$ angular resolution (regrid 2). The accretion rate $\dot{M}$ and bolometric luminosity $L_{\rm{bol}}$ roughly follow a $t^{-5/3}$ power law, indicated as the dashed line in the top panel. The magnetic flux after $t>10,000\, t_g$ quickly decreases to $\phi_{\rm{BH}}\approx30$ and is highly variable, indicating that flux eruptions are frequent.}
    \label{fig:KIa00diskthsli}
\end{figure*}

\begin{figure}
    \centering{}
	\includegraphics[width=\columnwidth]{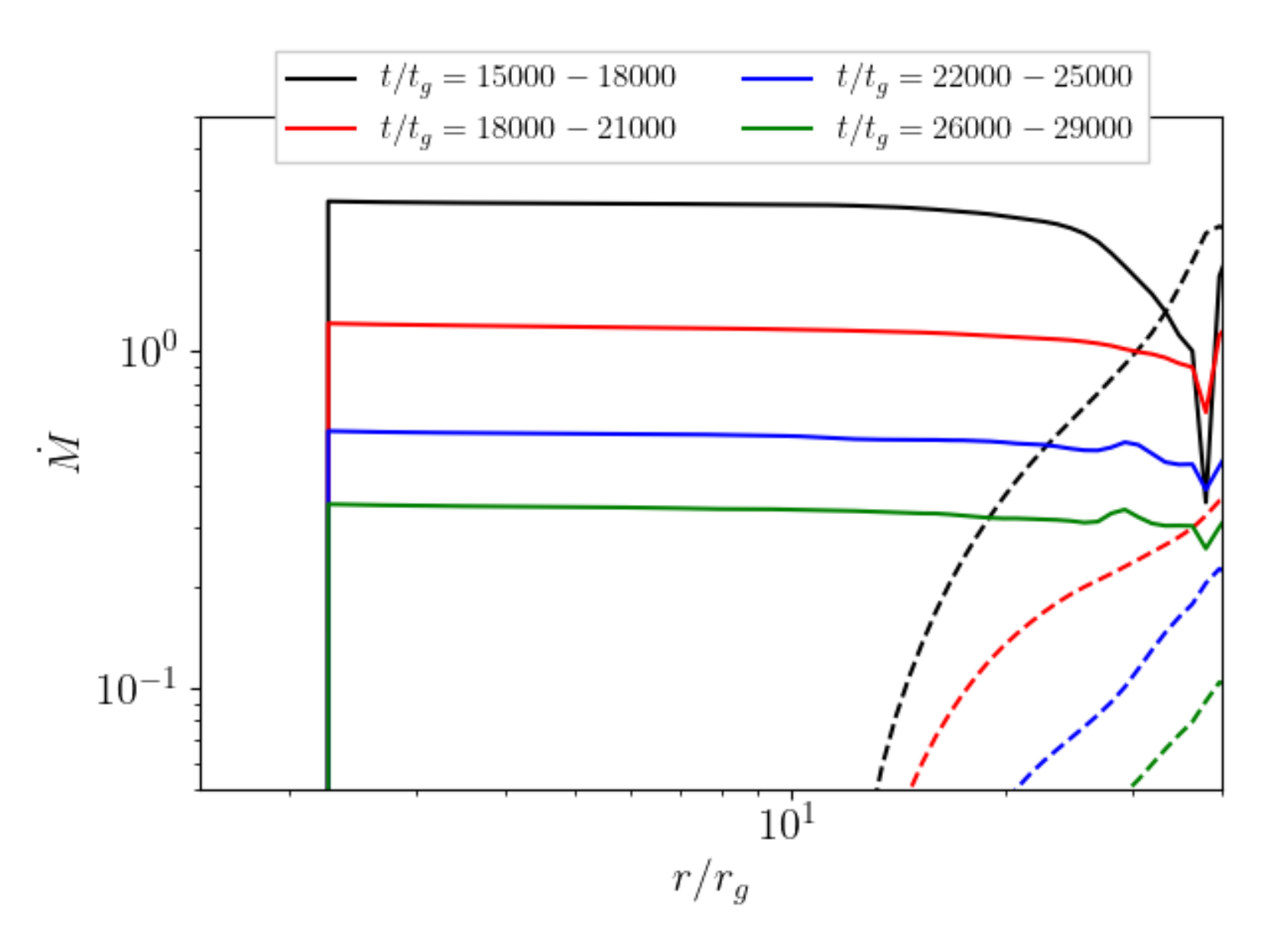}\\
	\includegraphics[width=\columnwidth]{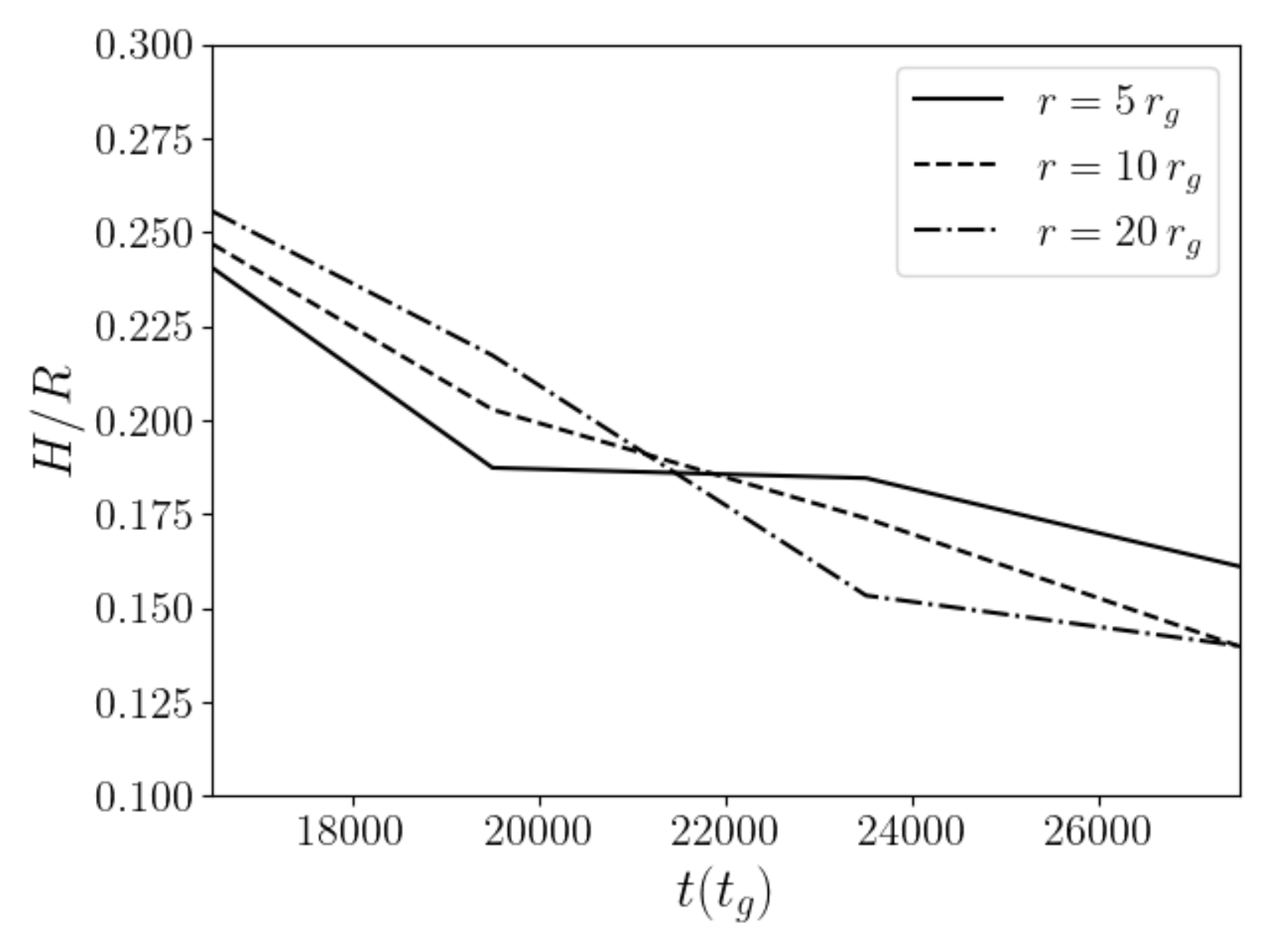}
    \caption{Here we present the time averaged radial profiles (top panel) for the accretion rate (solid lines) and mass outflow rate (dashed lines), and the density scale height as a function of time at selected radii (bottom panel), for \texttt{KIa00}. The time averaging range for each profile is indicated at the top of the panel. In the bottom panel, we show the density scale height throughout the evolution at $5\,r_g$ (solid line), $10\,r_g$ (dashed line), and $20\,r_g$ (dashed-dotted line).}
    \label{fig:KIa00scalars}
\end{figure}

\section{Results} \label{sec:results}

\subsection{Time Evolution}

\subsubsection{Zero Spin Model: \texttt{KIa00}} \label{sec:KIa00}

We show the full time evolution of the accretion rate $\dot{M}$, normalized magnetic flux at the BH horizon $\phi_{\rm{BH}}$, and the bolometric luminosity $L_{\rm{bol}}$ for \texttt{KIa00} in Figure \ref{fig:KIa00diskthsli}. Also shown are selected snapshots of the disk at the equatorial plane which span the full 3D evolution.

The accretion rate very closely follows the mass injection rate. As detailed in the previous section, we maintained a constant mass injection rate $\dot{M}_{\rm{inj}}$ until $t\geq 15,000\, t_g$, and the accretion rate $\dot{M}_{\rm{BH}}$ into the BH prior to this stage rises to a mean value of about $\dot{M}/\dot{M}_{\rm{Edd}} \approx 3$ and then remains nearly constant as a result. After we start to decrease the injection rate (see Equation \ref{eq:mdotinj}), the accretion rate similarly follows a $t^{-5/3}$ power law. By the end of the simulation, we sample an accretion state with $\dot{M}/\dot{M}_{\rm{Edd}} \approx 0.35$.

The magnetic flux grows rapidly in the initial 2D evolution from $t=0-10,000\, t_g$ and reaches a maximum value of $\phi_{\rm{BH}}\approx 55$ where it ceases to grow. Before the disk becomes MAD, it is the MRI which drives the angular momentum transport and hence the accretion. We find that the density weighted MRI quality factor integrated over all angles is $Q_\theta > 100$ during the entire evolution so MRI is well resolved in our simulations. After regridding to 3D, the magnetic flux escapes from the horizon in frequent flux eruption events which flow out radially and partially mix with the disk at larger radii. The most dramatic drop in flux occurs around $11,000\, t_g$ when $\phi_{\rm{BH}}$ decreases from $\approx55$ to $\approx 20$. As a result of the flux eruption events, the disk becomes highly non-axisymmetric throughout the entire evolution in 3D and the gas flows in primarily along dense spiral arms with much lower density gas filling the area in between, where the magnetic field is very strong. This behaviour is clearly shown in the top panels in Figure \ref{fig:KIa00diskthsli}. The net effect of the flux eruptions is that the magnetic flux saturates at a lower value than its initial peak, varying between $\phi_{\rm{BH}}\approx 20-40$ throughout. Flux eruptions are well known phenomena in MAD accretion disks (for instance, see \citealt{2008ApJ...677..317I,2012JPhCS.372a2040T}), but our disk appears to be more variable and have more severe flux eruptions in comparison to other models in the literature. This is likely due to the fact that the entire disk is MAD out to $R_{\rm{inj}}$, and we employ a unique disk initialization which drives the disk towards a large magnetic flux at all times. Gas flowing in from the injection radius is evidently able to efficiently transport magnetic flux inward and rapidly restores any lost magnetic flux at the horizon. The fact that $\phi_{\rm{BH}}$ is lower than the typical $\phi_{\rm{BH}}\sim 50$ seen in hot accretion flows in the MAD state (e.g. \citealt{2022MNRAS.511.3795N}) is likely because our systems are vertically thinner.

The luminosity behaves similarly to the accretion rate. As in the case of SANE, super-Eddington disks \citep{2016MNRAS.456.3929S,2019MNRAS.483..565C}, the luminosity is marginally super-Eddington when $\lambda = 2.8$ and maintains a value of $L_{\rm{bol}} \approx 2.8 L_{\rm{Edd}}$. At the end of the simulation when $\lambda = 0.35$, $L_{\rm{bol}}\approx 0.76 L_{\rm{Edd}}$.

We also consider how the simulation evolves in a time averaged sense by averaging the simulation data over $\phi$ and over chunks of $3000\, t_g$ in time. This allows us to sample the accretion state of the disk at various time intervals while ignoring the rapid variability; furthermore, the overall change in the disk properties over a given $3000\, t_g$ chunk of time is not substantial. As shown in Table \ref{tab:tab2}, the magnetic flux is stable across accretion rates with $\langle \phi_{\rm{BH}}\rangle \approx 32-33$. The luminosity does not follow precisely the same decline as $\lambda$, and $\langle L_{\rm{bol}}\rangle$ instead declines slightly more slowly. This has implications for the efficiency of the system, which we discuss later.

The net accretion rate as a function of radius is shown in the left panel of Figure \ref{fig:KIa00scalars}. We find that the profiles are nearly constant out to $r\approx 30\, r_g$ with a net outflow rate that is generally much less than the net inflow rate. Flat accretion rate profiles are typically seen in simulations where the initial conditions are an equilibrium magnetized torus. In such simulations, the gas flows in on the viscous accretion time and the disk is said to have reached inflow equilibrium up to the maximum radius where the net accretion rate is flat. The present simulations have a time dependent mass injection rate $\dot{M}_{\rm{inj}}$. Nevertheless, the profiles that we measure suggest that the disk is in a quasi inflow equilibrium state out to $r=30\, r_g$.

We measure the density scale height of the disk following Equation \ref{eq:eqHR} at radii $r=5,10,20\,r_g$ (bottom panel in Figure \ref{fig:KIa00scalars}, see also Table \ref{tab:tab2}). The disk exhibits substantial thinning at all radii as the accretion rate decreases. The initial scale height is $H/R\approx 0.25$ during the super-Eddington phase and increases slightly with radius. This is similar to the scale height of previous super-Eddington disks \citep{2015MNRAS.447...49S,2016MNRAS.456.3929S}. When the accretion rate drops to $\lambda = 0.35$, the scale height reaches $H/R\approx 0.15$ but the behavior as a function of radius is inverted, being smaller at $r=10,20\, r_g$ compared to $r=5\, r_g$. While the disk appears to thin somewhat as the accretion rate drops, it is clearly stable against runaway cooling and collapse. This is to be expected since the magnetic field provides pressure support and accounts for nearly half of the pressure in the disk. A similar result for thin MAD disks was also found by \citet{2018MNRAS.480.3547M,2022arXiv220103526L}.

Finally we come to the efficiencies $\eta$ of the various forms of energy outflow. Namely, we measure the total efficiency ($\eta_{\rm{tot}}$), the radiative efficiency ($\eta_{\rm{rad}}$), and the jet and wind efficiencies ($\eta_{\rm{jet}}$ and $\eta_{\rm{wind}}$, which both exclude radiation). We find that $\eta_{\rm{tot}}$ increases with decreasing accretion rate. We also observe the same trend in terms of $\eta_{\rm{rad}}$, which is initially only $5.7\%$ when the disk is super-Eddington and increases to $12.2\%$ when $\lambda=0.35$. We interpret this change as being related to the weaker advection of radiation as the disk scale-height and optical depth changes. For instance, the shrinking of the outer disk at $r\geq 10\,r_g$ to a smaller scale height than the inner disk suggests that radiative cooling is acting to remove thermal energy from the disk before it is advected across the BH horizon. Ultimately, this leads to a larger escaping radiant luminosity, relative to the accreted rest mass energy, and an increase in efficiency as $\lambda$ decreases. At the same time, $\eta_{\rm{jet}}$ increases from $1.9\%$ to $3.4\%$ while $\eta_{\rm{wind}}$ decreases from $1.1\%$ to $0.3\%$. This indicates that the mechanical component of the outflow energy (jet and wind) tilts from the wind to the jet as $\lambda$ decreases. Indeed, as we discuss later, the outflow velocity near $30\, r_g$ increases as the simulation evolves. See Table \ref{tab:tab2} for tabulated efficiencies as a function of time and accretion rate. 

Focusing on the sub-Eddington evolution, a total efficiency of $\eta_{\rm{tot}}\approx 15.1-15.9\%$ is more than three times the NT thin disk efficiency. Additionally, a radiative efficiency of $\eta_{\rm{rad}}\approx 11.6-12.2\%$ is more than two times the NT thin disk efficiency. Both metrics for the extracted energy show substantial differences from the theoretical value expected for a NT thin disk. We note that other authors have demonstrated deviations from the NT thin disk model in GRRMHD for MAD disks \citep{2021ApJ...919L..20D,2022arXiv220103526L}. GRMHD simulations which utilized a cooling function have also found significant deviations for MAD \citep{2016MNRAS.462..636A} and non-MAD \citep{2021ApJ...922..270K} accretion disks. The additional radiation that we find here may be provided by the binding energy of gas plunging into the BH. In particular, low angular momentum gas will have more binding energy available for dissipation. We find that the disk becomes significantly sub-Keplerian at $r<R_{\rm{inj}}$ and $u_\phi$ is nearly half the Keplerian angular momentum as it crosses the horizon, which suggests that this could be a contributing factor to the deviations we measure. There is also the matter of significant dissipation at radii inside the ISCO (which is also present in our simulations), which is not predicted in the NT model. We note that the inclusion of a two-temperature plasma at low accretion rates may significantly reduce emission. For example, \citet{2020ApJ...904..117K} find a coronal luminosity reduction of $\sim 1/2$ when contrasting a single temperature plasma with a two temperature plasma.

\begin{table*}
    \centering
    \begin{tabular}{ c c c c c c c c c c }
        \hline
        \hline 
        Model & Time & $\lambda$ & $\langle \phi_{\rm{BH}}\rangle$ & $\langle L_{\rm{bol}} \rangle$ & $(H/R)_{r=5\,r_g}$ & $\eta_{\rm{tot}}$ & $\eta_{\rm{rad}}$ & $\eta_{\rm{jet}}$ & $\eta_{\rm{wind}}$ \\
               & ($10^3 \,t_g$) & & &$(L_{\rm{Edd}})$  & & & & & \\
        \hline
        
        \texttt{KIa00} & 15-18 & 2.8 & 33 & 2.80 & 0.24 & 8.7\%  & 5.7\% & 1.9\% & 1.1\% \\
                       & 18-21 & 1.2 & 33 & 1.88 & 0.19 & 11.5\%  & 8.7\% & 2.5\% & 0.3\% \\
                       & 22-25 & 0.58 & 32 & 1.19 & 0.18 & 15.1\%  & 11.6\% & 3.2\% & 0.3\% \\
                       & 26-29 & 0.35 & 32 & 0.76 & 0.16 & 15.9\%  & 12.2\% & 3.4\% & 0.3\% \\ 
        \hline
        NT ($a_*=0$)   & - & - & - & - & - & 5.70\% & 5.70\% & - & - \\      
        \hline
         & & & & & & & & &\\
        \texttt{KIa09} & 14-17 & 2.2 & 36 & 6.70 & 0.18 & 88.0\%  & 48.2\% & 38.7\% & 0.6\% \\
                       & 20-23 & 0.76 & 33 & 3.68 & 0.15 & 100.1\%  & 75.1\% & 25.8\% & 0.6\% \\
                       & 23-26 & 0.44 & 29 & 2.21 & 0.15 & 98.3\%  & 78.8\% & 18.9\% & 0.6\% \\
                       & 28-31 & 0.31 & 26 & 0.99 & 0.13 & 68.3\%  & 50.2\% & 17.7\% & 0.5\% \\ 
                       & 31-34 & 0.29 & 27 & 0.85 & 0.12 & 64.3\%  & 45.4\% & 18.4\% & 0.4\% \\ 
        \hline
        NT ($a_*=0.9$)   & - & - & - & - & - & 15.58\% & 15.58\% & - & - \\          
    \hline
    \end{tabular}
    \caption{Tabulated values of the time- and $\phi$-averaged magnetic flux at the BH horizon ($\langle \phi_{\rm{BH}}\rangle)$, bolometric luminosity ($\langle L_{\rm{bol}}\rangle $), disk scale height ($H/R$), total efficiency ($\eta_{\rm{tot}}$), radiative efficiency ($\eta_{\rm{rad}}$), jet efficiency ($\eta_{\rm{jet}}$), and wind efficiency($\eta_{\rm{wind}}$), for each model as a function of time and Eddington-scaled accretion rate. We also list the corresponding NT thin disk efficiency for comparison. Note that the time range for time-averaging is shown in units of $10^3 \, t_g$ and $\lambda$ is the average accretion rate in units of $\dot{M}_{\rm{Edd}}$ during the given time range. The disk scale height presented here is computed at $r=5\, r_g$.}
    \label{tab:tab2}
\end{table*}

\begin{figure*}
    \centering{}
	\includegraphics[width=\textwidth]{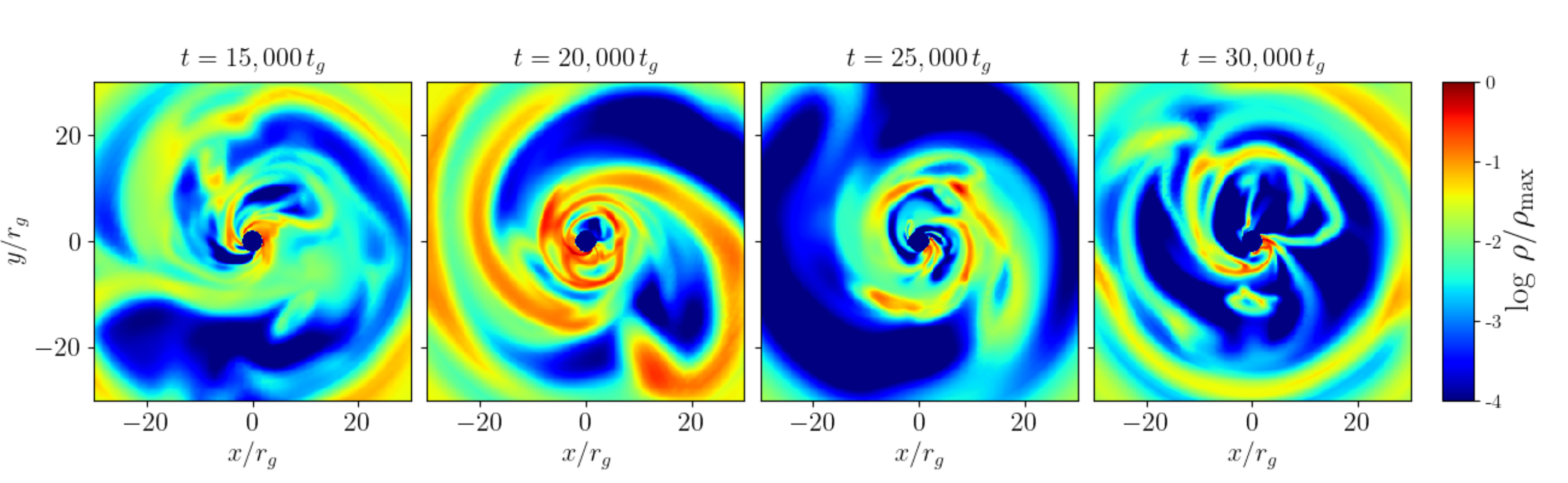}\\
	\includegraphics[width=\textwidth]{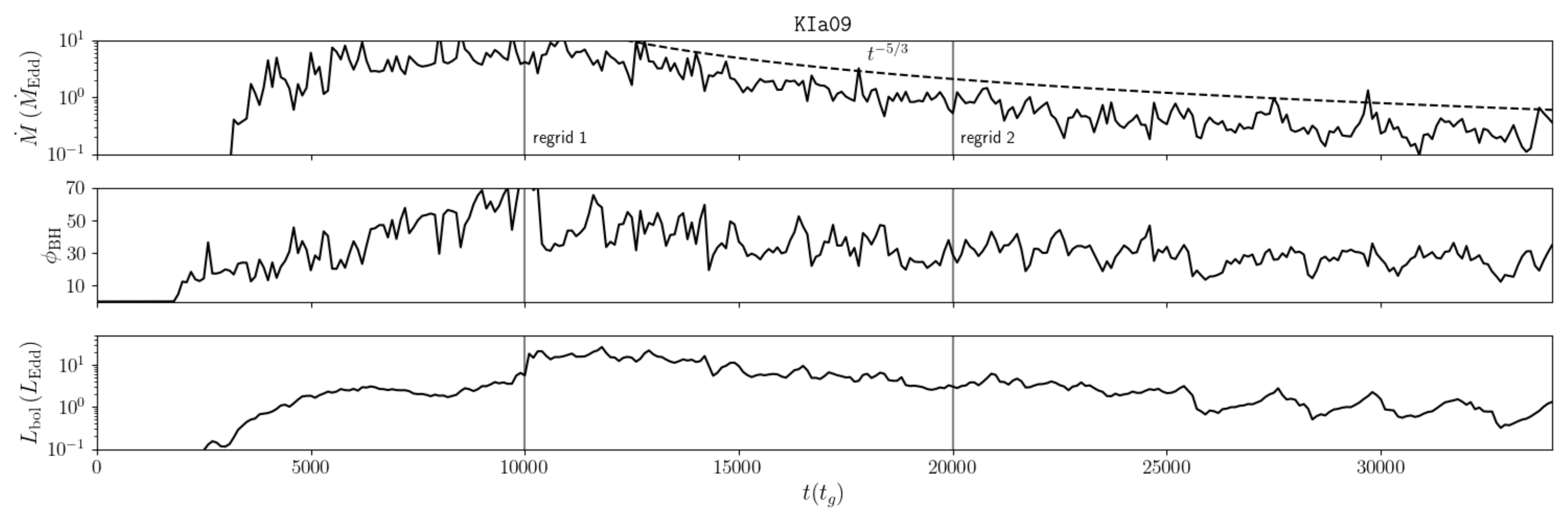}
    \caption{The same as Figure \ref{fig:KIa00diskthsli} but for \texttt{KIa09}. The simulation history in the bottom panels is similar to that found for \texttt{KIa00} but the luminosity is higher because of the contribution from a relativistic jet. Also note that the injection rate starts to decline at $t=10,000\, t_g$.}
    \label{fig:KIa09diskthsli}
\end{figure*}

\begin{figure}
    \centering{}
	\includegraphics[width=\columnwidth]{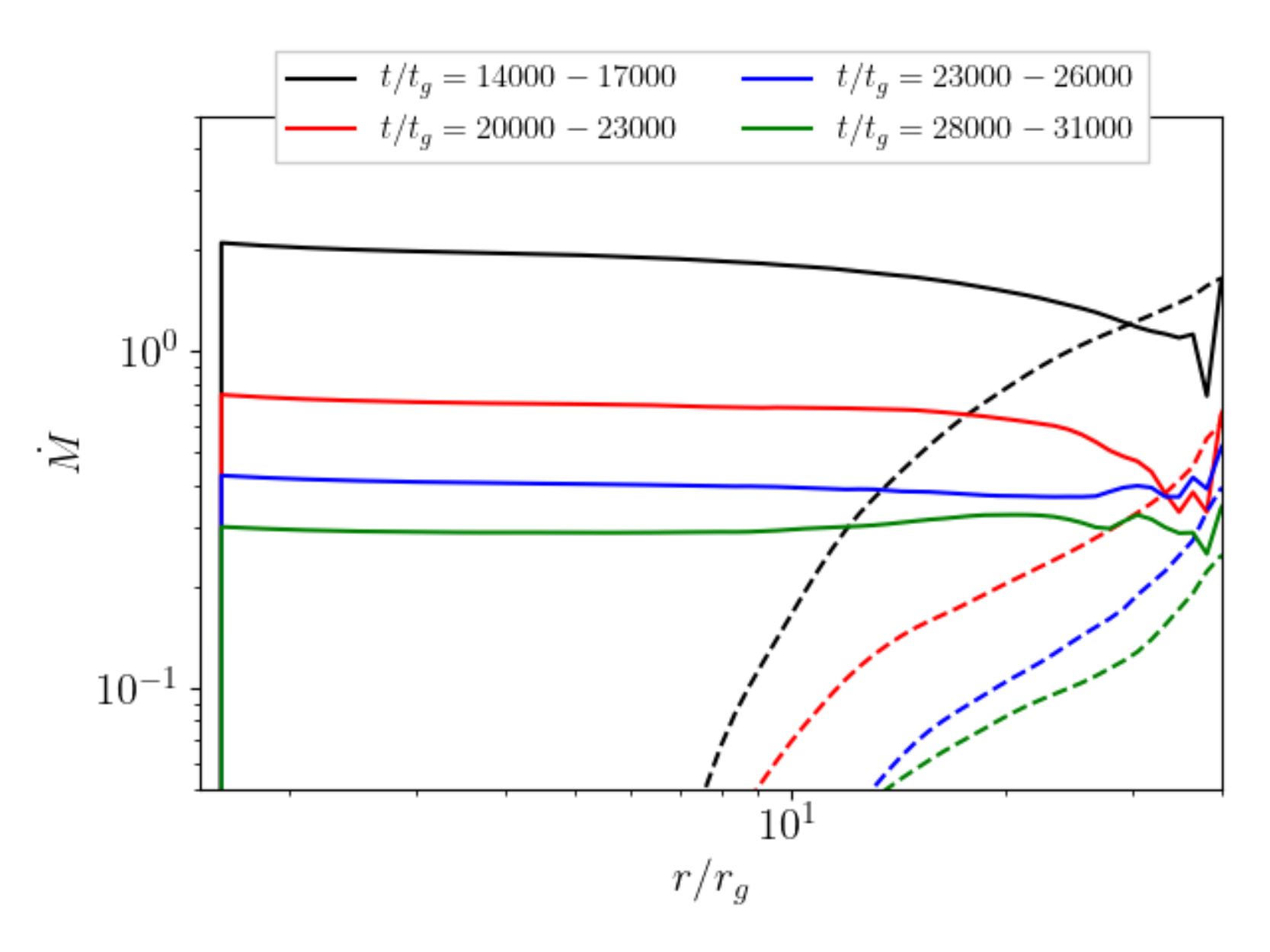}\\
	\includegraphics[width=\columnwidth]{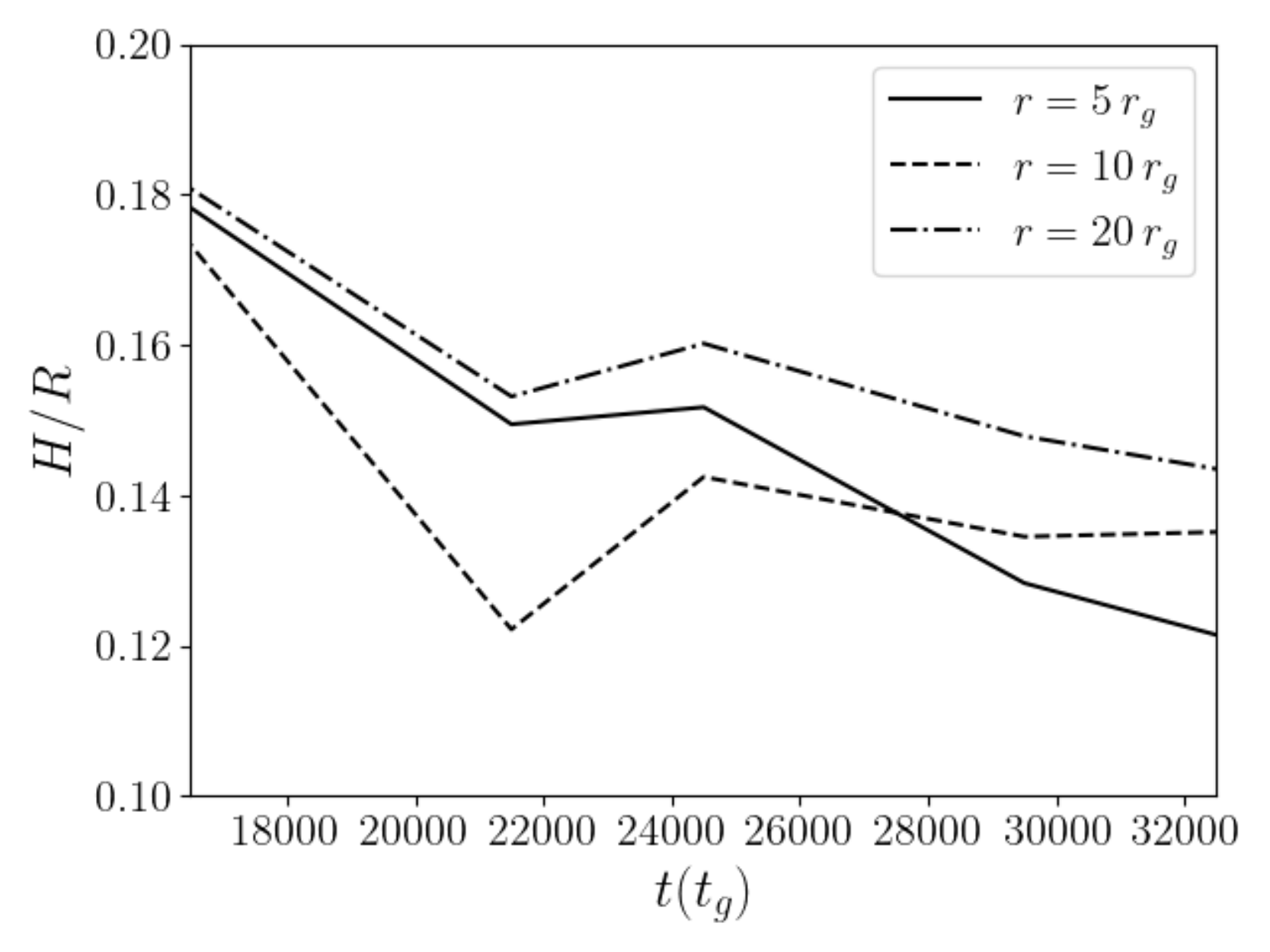}
    \caption{The same as Figure \ref{fig:KIa00scalars} but for \texttt{KIa09}.}
    \label{fig:KIa09scalars}
\end{figure}

\subsubsection{High Spin Model: \texttt{KIa09}} \label{sec:KIa09}

As demonstrated in Figure \ref{fig:KIa09diskthsli}, the evolution of \texttt{KIa09} is quite similar to \texttt{KIa00} in several respects. The accretion rate after we begin to decrease the injection rate (at $t>10,000\, t_g$) closely tracks the $t^{-5/3}$ behaviour of the injection rate, and the accretion rate decreases to $\approx0.29\dot{M}_{\rm{Edd}}$ by the end of the simulation. The magnetic flux rises to an initial saturation value of $\phi_{\rm{BH}}\approx 55$ but then declines during the 3D evolution before leveling off at $\phi_{\rm{BH}}\approx20-40$. The disk has frequent flux eruptions, and the BH again primarily accretes gas through dense spiral arms. We note that the initial flux eruption after `regrid 1' occurs at $t\approx 10,500\, t_g$ and is both earlier and far less dramatic than the first flux eruption in \texttt{KIa00} after `regrid 1'. The time averaged magnetic flux (Table \ref{tab:tab2}) declines throughout the simulation starting at $\langle \phi_{\rm{BH}}\rangle \approx36$ when $\lambda =2.2$ but it declines to $\langle \phi_{\rm{BH}}\rangle \approx26-27$ by the time $ \lambda  \leq 0.31$. This suggests that magnetic flux is advected less efficiently at lower accretion rates in \texttt{KIa09}, perhaps due to the lower $H/R$ that we measure. This is the opposite of \texttt{KIa00}, which showed essentially no change in magnetic flux by the end of the simulation; however, we note that $H/R$ was also generally larger in \texttt{KIa00}.

The accretion rate as a function of radius in \texttt{KIa09} also behaves similarly to that in \texttt{KIa00} (compare the top panels of Figures \ref{fig:KIa00scalars} and \ref{fig:KIa09scalars}), and we observe profiles which are nearly constant out to $r\approx 30 \, r_g$. The net outflow rate of mass is again substantially lower than the mass inflow at small radii, but at $r=40\, r_g$ it is nearly equal to the mass inflow. This is likely due to the BZ process, which leads to a substantially higher jet power than in the non-spinning case.

The two primary differences between \texttt{KIa00} and \texttt{KIa09} are that (i) the disk starts out noticeably thinner in \texttt{KIa09} (bottom panel in Figure \ref{fig:KIa09scalars}), and (ii) the total efficiency, jet efficiency, and radiative efficiency are all substantially higher in \texttt{KIa09} (Table \ref{tab:tab2}). Both of these differences come about because of the addition of a spinning BH and the fact that the disk is MAD. For example, the theoretical energy extraction via the BZ power assuming $a_*=0.9$ and $\phi_{\rm{BH}}=30$ in Equation \ref{equ:etaBZ} gives an efficiency due to the BZ process of $\eta_{\rm{BZ}}\approx 31\%$. This extracted energy powers a jet which supplies significant pressure and compresses the disk while simultaneously producing radiation and driving a more energetic outflow in the form of a relativistic outflow. This extracted energy is in addition to any energy extracted from the disk and comes instead from the BH spin. We note that our total efficiency is substantially higher than $\eta_{\rm{BZ}}$, but efficiencies larger than $\eta_{\rm{BZ}}$ are not uncommon for MAD disks \citep{2012JPhCS.372a2040T,2012MNRAS.423.3083M,2022MNRAS.511.3795N,2022arXiv220103526L}.

The inner disk gets compressed due to the jet, and we observe a peak scale height of $H/R\approx 0.18$ (compared to $H/R\approx 0.25$ in \texttt{KIa00}, this is similar to the results reported by \citealt{2022MNRAS.511.3795N}). Unlike \texttt{KIa00}, the disk does not have an inverted $H/R(r)$ profile as the accretion rate approaches $\approx 0.3\dot{M}_{\rm{Edd}}$ and instead the outer disk reaches $H/R(r=20\, r_g)\approx 0.14$ and the inner disk reaches $H/R(r=5\, r_g)\approx 0.12$.

We find that $\eta_{\rm{tot}}$ and $\eta_{\rm{jet}}$ have a downward trend as the accretion rate drops. Notably, $\eta_{\rm{tot}}\approx 88.0\%$ when $\lambda = 2.2$ and increases to $\eta_{\rm{tot}}\approx100\%$ when $\lambda =0.44- 0.76$. However, as the accretion rate declines further to $\lambda = 0.29$, $\eta_{\rm{tot}}\approx 64.3\%$. On the other hand, $\eta_{\rm{jet}}$ is highest ($\approx 38.7$\%) when the accretion rate is super-Eddington and declines with $\lambda$, reaching only $\approx 18.4\%$ when $\lambda = 0.29$. The wind efficiency remains fairly stable at $\eta_{\rm{wind}} \approx 0.4-0.6\%$ throughout the entire evolution. See Table \ref{tab:tab2} for tabulated efficiencies in \texttt{KIa09}. Here we again find a large deviation from the NT thin disk, with the total efficiency exceeding NT by a factor of nearly $4$ and the radiative efficiency exceeding NT by a factor of $3$ during the lowest accretion states. 

We note that the numerical ceiling implemented in our simulations which requires $\sigma_M \leq 100$ likely impacts the radiative efficiency in \texttt{KIa09}. Mass is injected into the jet to maintain $\sigma_M \leq 100$. This ultimately leads to some dependence on the numerical floors in the radiation field in the jet since the gas in this region is both hot and magnetized and will thus produce radiation. This uncertainty is difficult to quantify even where $1<\sigma_M<100$ since the jet is optically thin and the radiation from regions effected by floors will spread throughout the jet (e.g. \citealt{2017MNRAS.467.2241M,2019MNRAS.486.2873C}). We performed a rerun of the 2D portion of \texttt{KIa09} to quantify the impact of the $\sigma_M$ ceiling on our results and find that setting a ceiling of $\sigma_M\leq20$ decreases the radiative efficiency as $L_{\rm{bol}}$ is decreased by roughly a factor of 2. The $\sigma_M$ ceiling is less of a concern in \texttt{KIa00} as the region where the numerical ceiling is applied is much smaller in extent (see Figure \ref{fig:sigma_comparison} in Appendix \ref{sec:appB}).

\subsection{Outflow Properties and Jet Structure}

\begin{figure*}
    \centering{}
	\includegraphics[width=\textwidth]{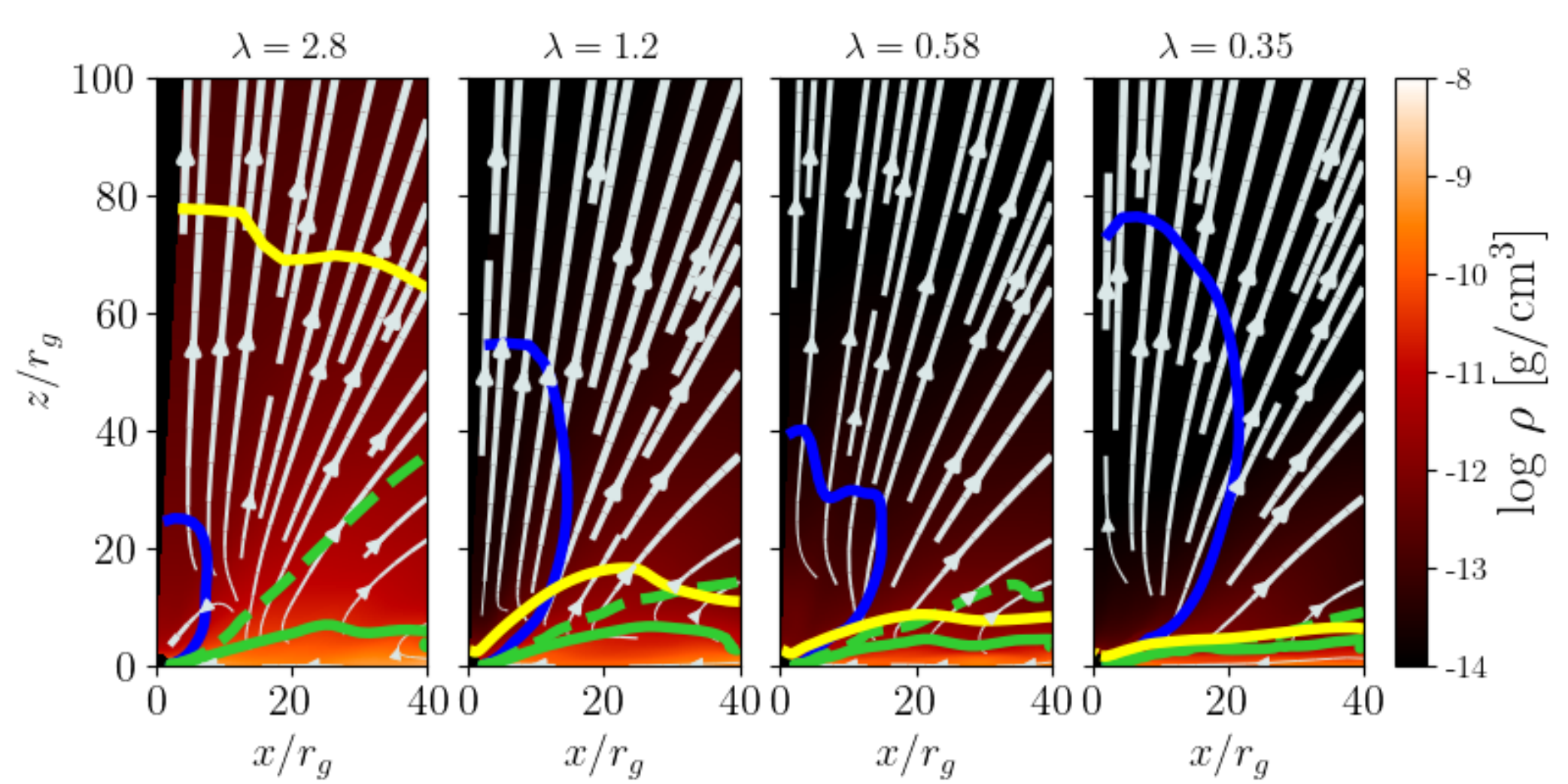}
    \caption{Here we show snapshots of \texttt{KIa00} for time-averages over $3000\, t_g$ and $\phi$-averaged. The time averaged accretion rate ($\lambda$) is indicated above each panel. The gas density (colors), fluid velocity (streamlines), jet/wind boundary (${\rm{Be}}=0.05$, dashed green line), wind/disk boundary (${\rm{Be}}=0$, green line), electron scattering photosphere ($\tau_{\rm{es}}=1$, yellow line), and magnetized jet boundary ($\sigma_M=1$, blue line) are indicated in each panel.}
    \label{fig:diskphiavg_KIa00}
\end{figure*}

\begin{figure*}
    \centering{}
	\includegraphics[width=\textwidth]{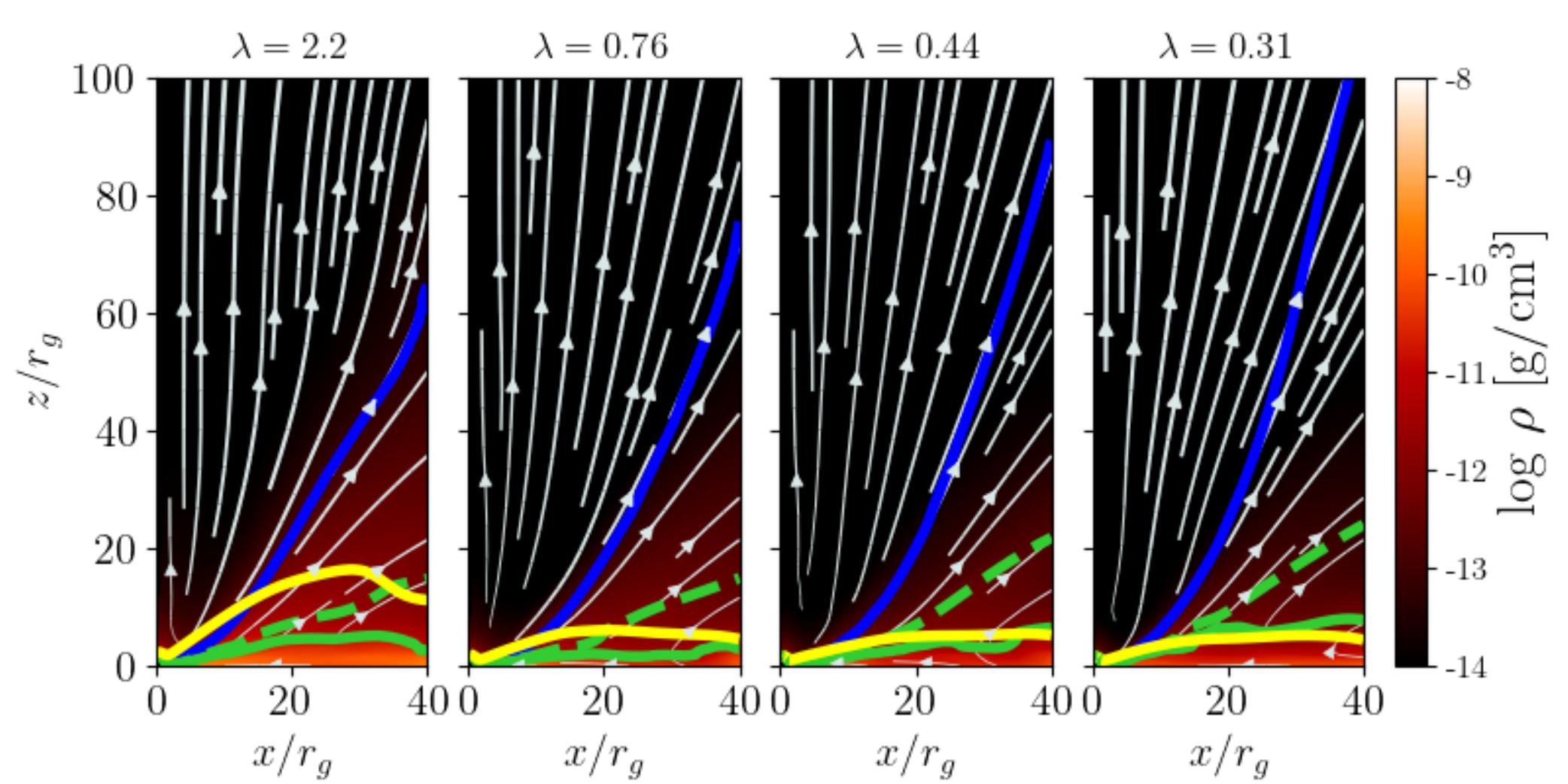}
    \caption{The same as Figure \ref{fig:diskphiavg_KIa00} but for model \texttt{KIa09}.  As the accretion rate declines, the jet power declines substantially and the jet width declines.}
    \label{fig:diskphiavg_KIa09}
\end{figure*}

\begin{figure*}
    \centering{}
	\includegraphics[width=0.98\columnwidth]{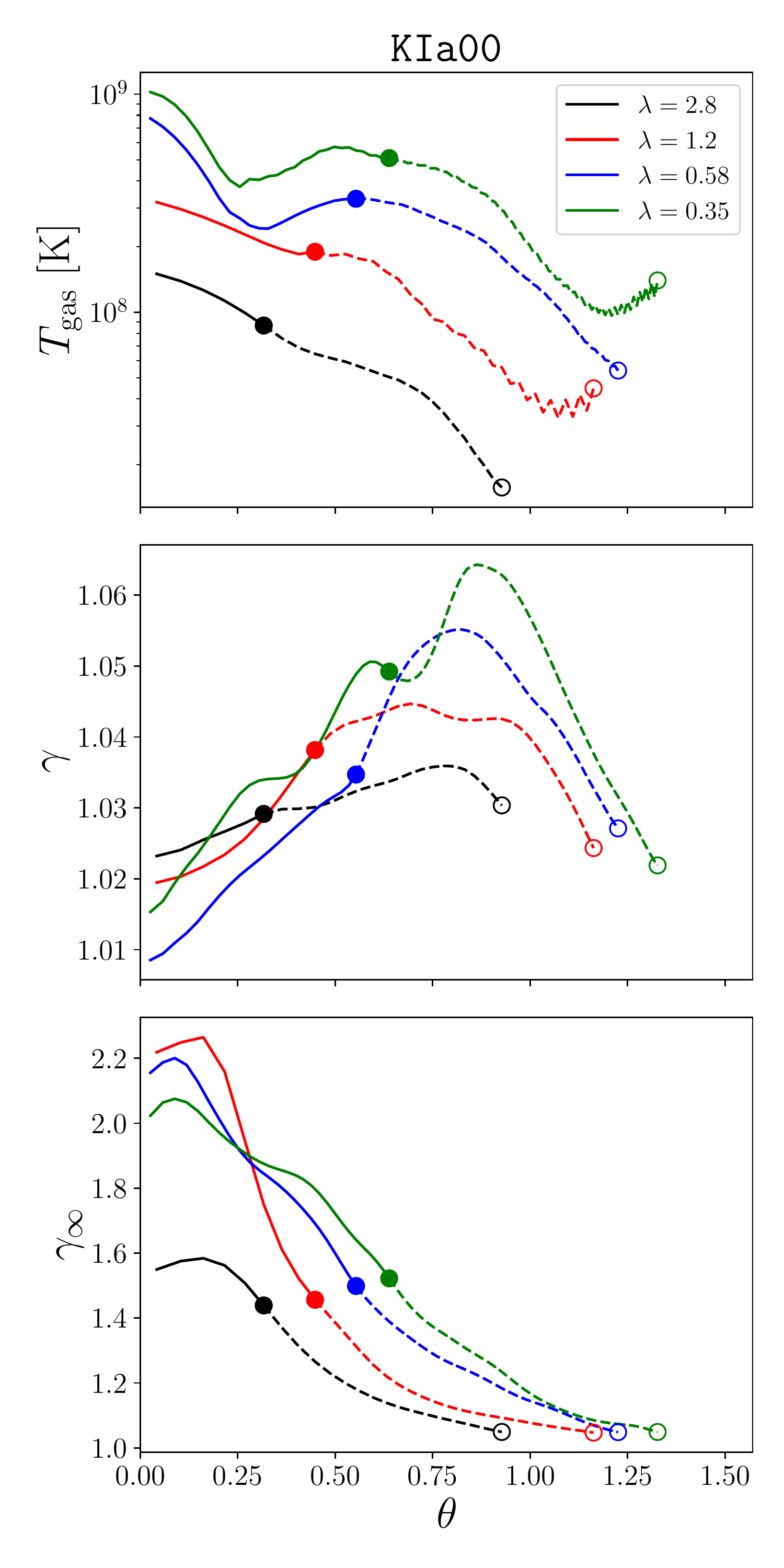}
	\includegraphics[width=0.98\columnwidth]{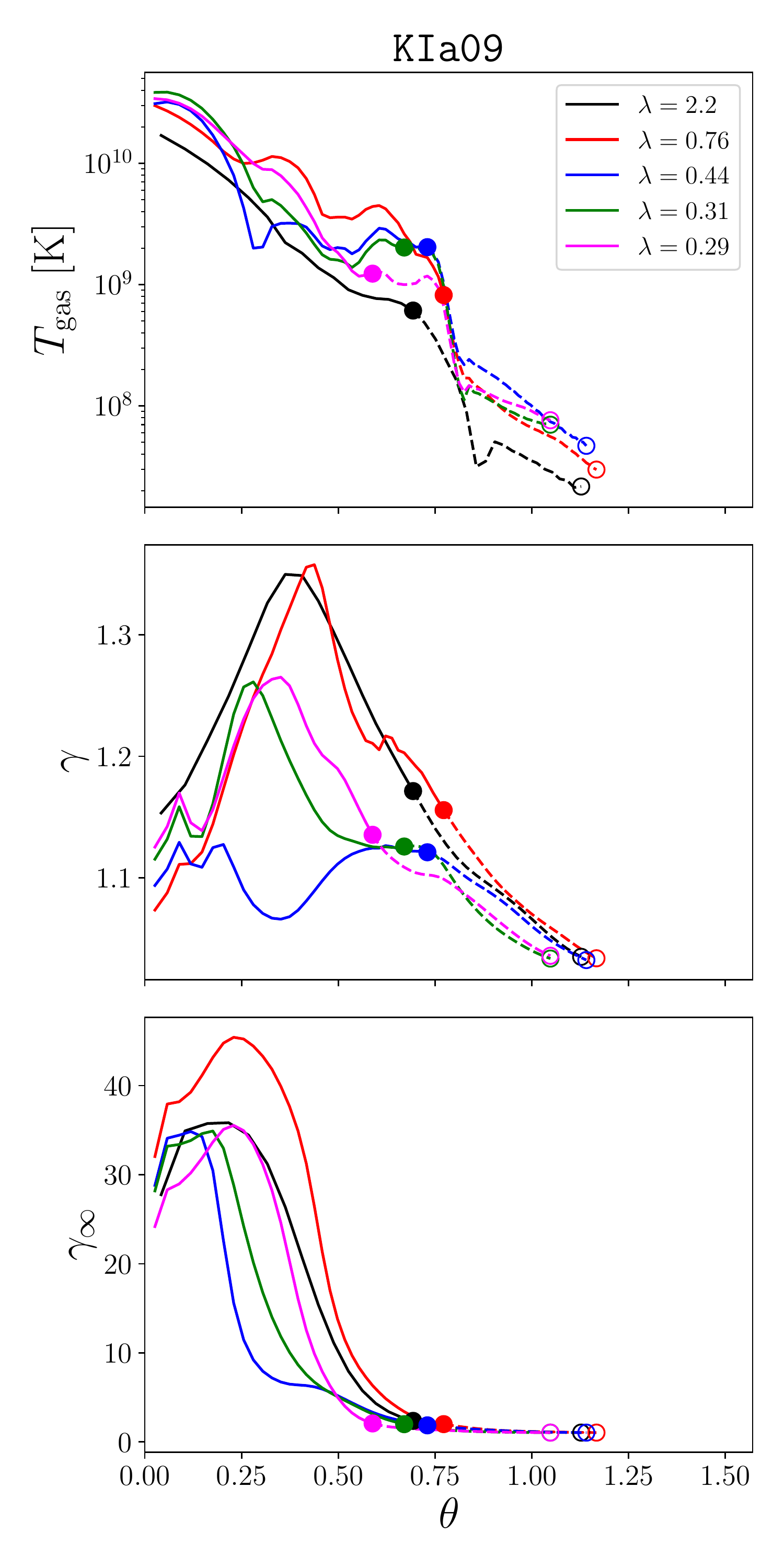}
    \caption{Here we show angular profiles of the gas temperature (top), lorentz factor (middle), and asymptotic lorentz factor (bottom) in the jet (${\rm{Be}}\geq 0.05$) for \texttt{KIa00} (left) and \texttt{KIa09} (right) as a function of polar angle $\theta$ at $r=30\, r_g$. Note that we have symmetrized the data across $\theta=\pi/2$ such that quantities shown are an average of the top and bottom jet. The solid line is for regions where $\sigma_M > 1$ while the dashed line signifies where $\sigma_M < 1$ and ${\rm{Be}}\geq0.05$. We mark the location where $\sigma_M = 1$ (solid circle) as well as where ${\rm{Be}} = 0.05$ (open circle). The temperature of the outflow in \texttt{KIa00} systematically increase with decreasing accretion rate. Both $\gamma$ and the asymptotic lorentz factor $\gamma_\infty$ also generally increase with decreasing $\lambda$ with the exception of the $\sigma_M>1$ region (especially near the poles). \texttt{KIa09} does not show significant differences in the jet temperature, nor do $\gamma$ or $\gamma_\infty$ show clear trends in their behaviour as $\lambda$ decreases. However, $\gamma$ during the $3000\,t_g$ period that we average over where $\lambda=0.44$ is much smaller than at other points in the simulation due to a prolonged drop in jet power, as we discuss in the text. The peak asymptotic lorentz factor suggests the outflow in \texttt{KIa00} may accelerate to mildly relativistic outflow with $\gamma_\infty\sim1.6-2.2$ while the outflow in \texttt{KIa09} may reach highly relativistic speeds of $\gamma_\infty\gtrsim30$ at all mass accretion rates $\lambda$ that we have sampled.}
    \label{fig:TandGamma}
\end{figure*}

\begin{figure}
    \centering{}
	\includegraphics[width=\columnwidth]{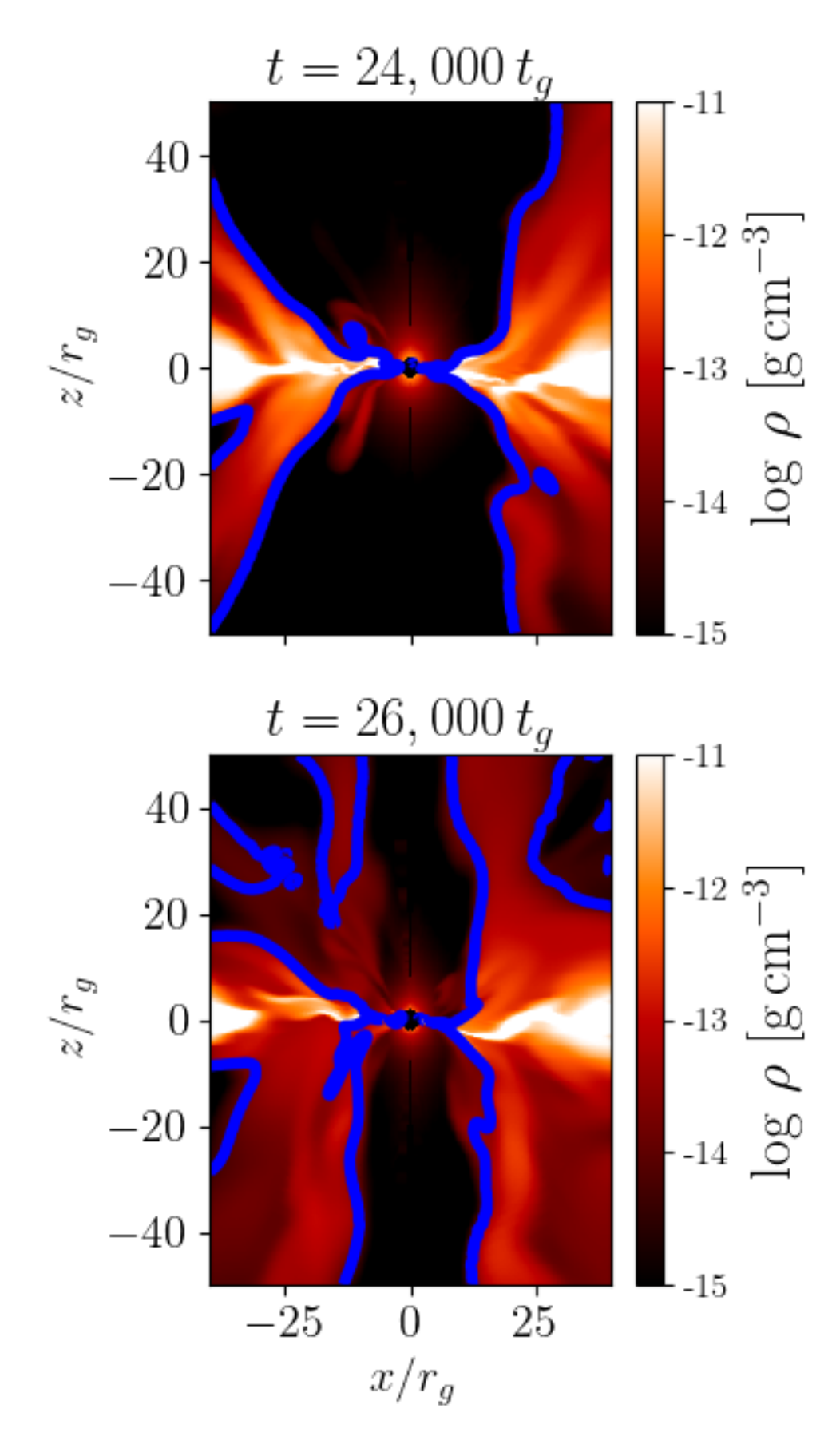}
    \caption{Here we show snapshots of the gas density (colors) and magnetized jet boundary $\sigma_M=1$ (blue line) for \texttt{KIa09} between $t=23,000-26,000\,t_g$ when the jet power is high (top panel) and when the jet power is low (bottom panel). At $t=24,000\,t_g$ (top panel), the normalized magnetic flux at the BH horizon is $\phi_{\rm{BH}}\approx35$ and the instantaneous jet power is $\eta_{\rm{jet}}\approx22\%$. Here a powerful jet creates a low density funnel where the magnetic pressure dominates. At $t=26,000\,t_g$ (bottom panel), $\phi_{\rm{BH}}\approx15$ and  $\eta_{\rm{jet}}\approx3\%$. The polar region in this low jet power state is filled in with higher density gas with $\sigma_M<1$ and the overall width of the jet narrows. The low jet power state in the bottom panel is prolonged for $t=23,000-26,000\,t_g$ ($\lambda=0.44$) and leads to a lower jet velocity in a time averaged sense.}
    \label{fig:KIa09jet}
\end{figure}

\begin{figure}
    \centering{}
	\includegraphics[width=\columnwidth]{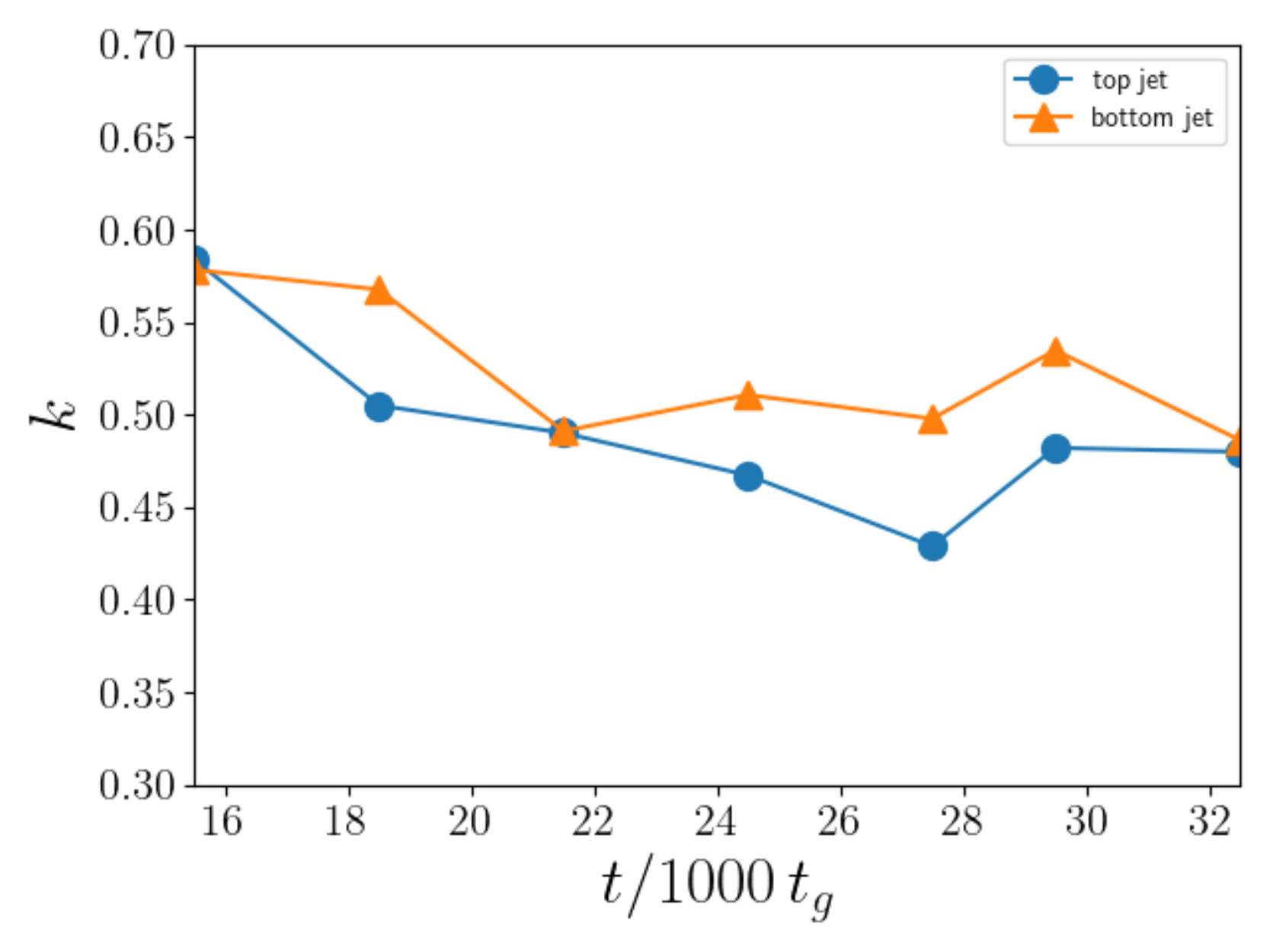}
    \caption{Here we show fits to the jet radius as a function of jet height, $w\propto z^k$, at different times during the evolution of \texttt{KIa09} by fitting the $\sigma_M=1$ profile over radii in the range $r=10-100 \, r_g$ for both the top and bottom jet. The power-law index $k$ appears to be slightly larger during the super-Eddington phase of the simulation, which coincides with a higher jet power.}
    \label{fig:jetedge}
\end{figure}

Figures \ref{fig:diskphiavg_KIa00} and \ref{fig:diskphiavg_KIa09} show the disk, wind and jet structure as a function of accretion rate for the two simulations. 

For \texttt{KIa00}, we find that at high accretion rates ($\lambda\approx 2.8$), the photosphere is extended in the polar region, extending to nearly $80\, r_g$. As the accretion rate decreases to $\lambda\sim 1$, the polar region becomes optically thin which allows for radiative acceleration close to the BH horizon. The overall scale of the wind region shrinks considerably as the accretion rate drops, which agrees with the fact that the measured wind efficiency declines from 1.1\% to 0.3\% by the time $\lambda = 0.35$ while the jet efficiency increases from 1.9\% to 3.4\%. The enhanced energy deposited in the jet is due to the shrinking of the photosphere, which is nearly coincident with the disk boundary for sub-Eddington values of $\lambda$. Radiative coupling allows greater acceleration as the optical depth drops, which leads to larger total energy in the gas streaming away from the disk surface. The ${\rm{Be}}=0$ surface,  which is the disk-wind boundary, also reflects our results for the density scale height (Figure \ref{fig:KIa00scalars}), and the disk shows considerable changes in height as the accretion rate declines.

We present angular profiles in $\theta$ of the gas temperature, lorentz factor, and asymptotic lorentz factor $\gamma_\infty$ as measured at $r=30\, r_g$ in Figure \ref{fig:TandGamma}. We compute the asymptotic lorentz factor from the Bernoulli (Equation \ref{eq:Bernoulli}) by assuming all other forms of energy are converted to kinetic energy far from the BH. Since $u_t\approx -\gamma$ at $r\gg r_g$, we compute the asymptotic lorentz factor using: 
\begin{equation}
    \gamma_\infty = {\rm{Be}} + 1.
\end{equation}
The initial gas temperature at $\lambda=2.8$ reaches $T_{\rm{gas}}\approx 10^7-10^8$ K, but when $\lambda=0.35$, the temperature has increased to nearly $T_{\rm{gas}}\approx 10^8-10^9$ K. The outflowing gas in \texttt{KIa00} achieves speeds in excess of $0.23c$ by $r=30\, r_g$, and the gas velocity increases monotonically as the accretion rate declines, with the exception of the $\sigma_M>1$ region which shows no clear trend across $\lambda$. We find that when $\lambda=2.8$, the peak outflow velocity is $\approx 0.23 c$, but when $\lambda=0.35$ it is $\approx0.33c$. The gas temperature in the jet increases as the accretion rate shrinks. The asymptotic velocity also shows a monotonic increase in the outflow velocity as $\lambda$ decreases aside from near the poles in the $\sigma_M >1$ region. Despite the lack of BZ power in \texttt{KIa00}, $\gamma_\infty$ implies the outflow may achieve a mildly relativistic asymptotic velocity of $\sim0.78-0.86c$.

In the case of \texttt{KIa09}, the pole is always optically thin down to the BH horizon for the accretion rates we study in this work. In addition, the jet features aren't strongly affected by the accretion rate. The gas temperature in the jet is generally $T_{\rm{gas}} > 10^{8}$ K, with the hottest gas residing in the $\sigma_M > 1$ region where $T_{\rm{gas}}\sim 10^9 - 10^{10}$ K. The jet achieves more relativistic velocities in this model compared to \texttt{KIa00} (see Figure \ref{fig:TandGamma}) with a peak of $v\approx0.65c$ when $\lambda = 0.76-2.2$ (where the jet power was observed to be highest). For $\lambda= 0.29-0.44$ the peak velocity reaches $v\approx0.6c$; however, there is a noticeable decline to $v\approx 0.4 c$ at $\lambda=0.44$ which we discuss in the following paragraph. The $\sigma_M=1$ boundary becomes visibly less wide over time. This agrees well with the shrinking jet power. The profiles of $\gamma_\infty$ illustrate definitively that the relativistic jet does not shut off in \texttt{KIa09}. Across all values of $\lambda$ that we sample, $\gamma_\infty$ reaches $\gtrsim 30$. Unlike in \texttt{KIa00}, there is no clear trend in $\gamma_\infty$, measured at $r=30\,r_g$, as $\lambda$ decreases for model \texttt{KIa09}. It is possible that interaction with the wind as the jet expands outward could modify the angular profiles of $\gamma_\infty$ as a function of radius and slow some of the outflow into a slower moving jet sheath (as in \citealt{2019MNRAS.490.2200C}), but we are restricted to measuring the outflow properties at small radii due to our implementation of a mass injection scheme.

When $\lambda=0.44$, the disk goes through two of its longest periods (at $t\approx 23,500\, t_g$ and $t\approx 25,500\, t_g $) where $\phi_{\rm{BH}}$ reaches a minimum of $14-19$, substantially lower than the mean of $\approx29$ (Table \ref{tab:tab2}). Here $\phi_{\rm{BH}}$ takes longer to recover ($\sim400\, t_g$ and $\sim700\, t_g$, respectively) than during other magnetic flux eruptions, where the magnetic flux only takes roughly $100-200\, t_g$ to increase back to its mean value. The rather extreme nature of the magnetic flux eruptions during this stage is well captured at $t=25,000\,t_g$ (3rd panel from the left in Figure \ref{fig:KIa09diskthsli}), where the accretion flow is essentially an inner disk surrounded by low density gas where magnetic islands are orbiting at $r\gtrsim 20\, r_g$. We show snapshots of the simulation when the jet power is high versus low in Figure \ref{fig:KIa09jet}. The $\sigma_M=1$ boundary is closer to the pole and the jet velocity is substantially lower when the jet power is weak. Indeed, the instantaneous jet efficiency at these points is $\eta_{\rm{jet}}\sim3-7\%$. In addition, when the jet power temporarily drops for extended periods of time, higher density gas launched by the disk wind fills in the polar regions and mass loads the jet. As a result, the velocity (in a time averaged sense) over the time range $t=23,000-26,000\,t_g$ is noticeably smaller with a maximum of $v\approx0.4c$ at $30\,r_g$. The profiles of $\gamma_\infty$ at $\lambda=0.44$ suggest that at low jet power states the highly relativistic component is not entirely shut off. Rather, a smaller fraction of the jet gets accelerated to $\gamma_\infty\sim30$ (between $\theta\sim 0-0.25$) while the mass loaded component at larger angles reaches only $\gamma_\infty \lesssim 6$.

Note that the $\sigma_M=1$ boundary does not extend to large radii in \texttt{KIa00} but closes off at $r\sim 20-80\, r_g$ (Fig.~\ref{fig:diskphiavg_KIa00}), whereas it extends to infinity in \texttt{KIa09} (Fig.~\ref{fig:diskphiavg_KIa09}). This is indicative of the lack of a magnetized relativistic jet in \texttt{KIa00}. Hence, we only measure the jet profile in \texttt{KIa09}. To measure the jet profile of \texttt{KIa09}, we treat the relativistic jet as the region with $\sigma_M > 1$ and take the polar boundary of the top/bottom jet to be where $\sigma_M = 1$. We assume that the profile takes the form $w\propto z^k$, where $w=r\sin\theta_{\rm{jet}}$ is the jet radius, $z=r\cos\theta_{\rm{jet}}$ is the jet height, and $k$ is a constant. 
We measure the jet profile between $r=10-100\,r_g$ using time averages of $\Delta t = 3,000 \, t_g$. This choice of radial limits is well motivated as the jet profile within this range is approximately linear in $\log z - \log w$ space. We perform a linear fit in $\log z - \log w$ space to obtain the constant $k$ as a function of time. As shown in Figure \ref{fig:jetedge}, the jet profile appears to be initially slightly less parabolic for super-Eddington accretion rates with $k\approx 0.58$ for both the top and bottom jet. As the accretion rate declines and the jet power decreases, the jet settles into a very nearly parabolic profile with $k\approx0.43-0.49$ for the top jet and $k\approx0.49-0.53$ for the bottom jet. The asymmetry between the top and bottom jet is not unusual, as other GRRMHD simulations have observed similar asymmetrical jets in BH accretion disks (e.g. see \citealt{2018MNRAS.480.3547M}).

These results reflect a similar trend to the recently reported results of \citet{2022MNRAS.511.3795N} who also found that the jet parameter $k$ decreased as the jet power decreased. However, our jet profile constants are slightly higher in general, e.g. their $a_*=0.9$ model had $k\approx0.43$. It is interesting that our measured jet profiles are similar to those of large scale AGN jets. For example, \citet{2020MNRAS.495.3576K} report jets with $k\approx$ 0.39 - 0.56 while \citet{2021A&A...647A..67B} find $k\approx$ 0.39-1.86. The jet profile of \texttt{KIa09} is comparable to 3C 120, which is a high state AGN whose Eddington ratio may be as high as $\lambda\approx 0.3$ \citep{2009ApJ...704.1689C} with a profile at small radii ($r<1$ parsec) with $k\approx 0.586\pm0.047$ \citep{2020MNRAS.495.3576K}. As was noted by \citet{2022MNRAS.511.3795N}, the observed jet profiles of AGN are derived from emission produced by the jet sheath, which likely does not coincide with $\sigma_M=1$. Thus, our jet profiles may be narrower than those inferred directly from observations. Ray traced radio images may better quantify jet profiles from GRRMHD simulations of thin disks, but this is beyond the scope of the present work.

Our results suggest that jets from a thin, highly magnetized accretion disk can explain the jet profile at small radii seen in some AGN. 
Our model with $\langle \phi_{\rm{BH}}\rangle \approx 30$ tends towards a slightly less magnetized disk than geometrically thick, MAD ADAF models, which have $\langle\phi_{\rm{BH}}\rangle \approx 50$ (e.g. \citealt{2022MNRAS.511.3795N}). In addition, our simulation differs from other works which have applied accretion disk simulations to study the structure of AGN jets (e.g. \citealt{2019MNRAS.490.2200C}) in that we do not have a large scale torus which may provide confinement at larger radii. Instead, all of the confinement happens due to the wind launched from the disk. 

An important point to note is that our simulation setup forces the system to be in the MAD state. It is suspected that thin disks in AGNs will allow magnetic flux to diffuse away from the BH, thereby preventing the high magnetic flux that we report in this work \citep{1994MNRAS.267..235L,1996ApJ...473..403H}. This is in conflict with the presence of jets in AGN with luminosities corresponding to the thin disk regime, which suggests that some of these AGN accretion disks are capable of advecting magnetic field towards the BH. For AGN that do maintain a significant magnetic flux, the total jet and wind power will be non-negligible and confinement similar to what we find in this work may be expected. Further simulations of steady state, thin, MAD accretion disks are warranted.

\subsection{Variability}

MAD accretion disks are by nature highly variable and have been cited as potential sources of quasi-periodic oscillations in AGN and other astrophysical sources with BH accretion as the central engine \citep{2012JPhCS.372a2040T,2014MNRAS.437.2744T}. In general, it would be ideal to study variability from physically motivated simulations such as presented in this work. As was noted by \citet{2016MNRAS.456.3929S}, while numerical simulations can provide detailed information on the radiation field, the accuracy of the radiation field is affected by the numerical schemes used and the treatment of radiation. In particular, the M1 closure scheme which is applied by \textsc{KORAL} assumes a gray approximation in the radiative transfer and a simple angular distribution of the radiation field \citep{2013MNRAS.429.3533S}. 

With those complications in mind, we attempt to measure the variability directly from the \textsc{KORAL} simulation. One would ideally measure the radiative flux at a large distance (i.e. at the outer boundary of the simulation domain) as these rays would be expected to reach a distant observer. This is not accurate though if the system is not in equilibrium out to the radius at which the flux is measured. If the radiation is decoupled from the gas the radiation field should show similar variability at each radius with a time shift due to the difference in the light crossing time. The present simulations are particularly complicated because although the jet and magnetized wind are both optically thin, the disk properties are changing on roughly the viscous time so, by construction, the system is never in equilibrium.

\begin{figure}
    \centering{}
	\includegraphics[width=\columnwidth]{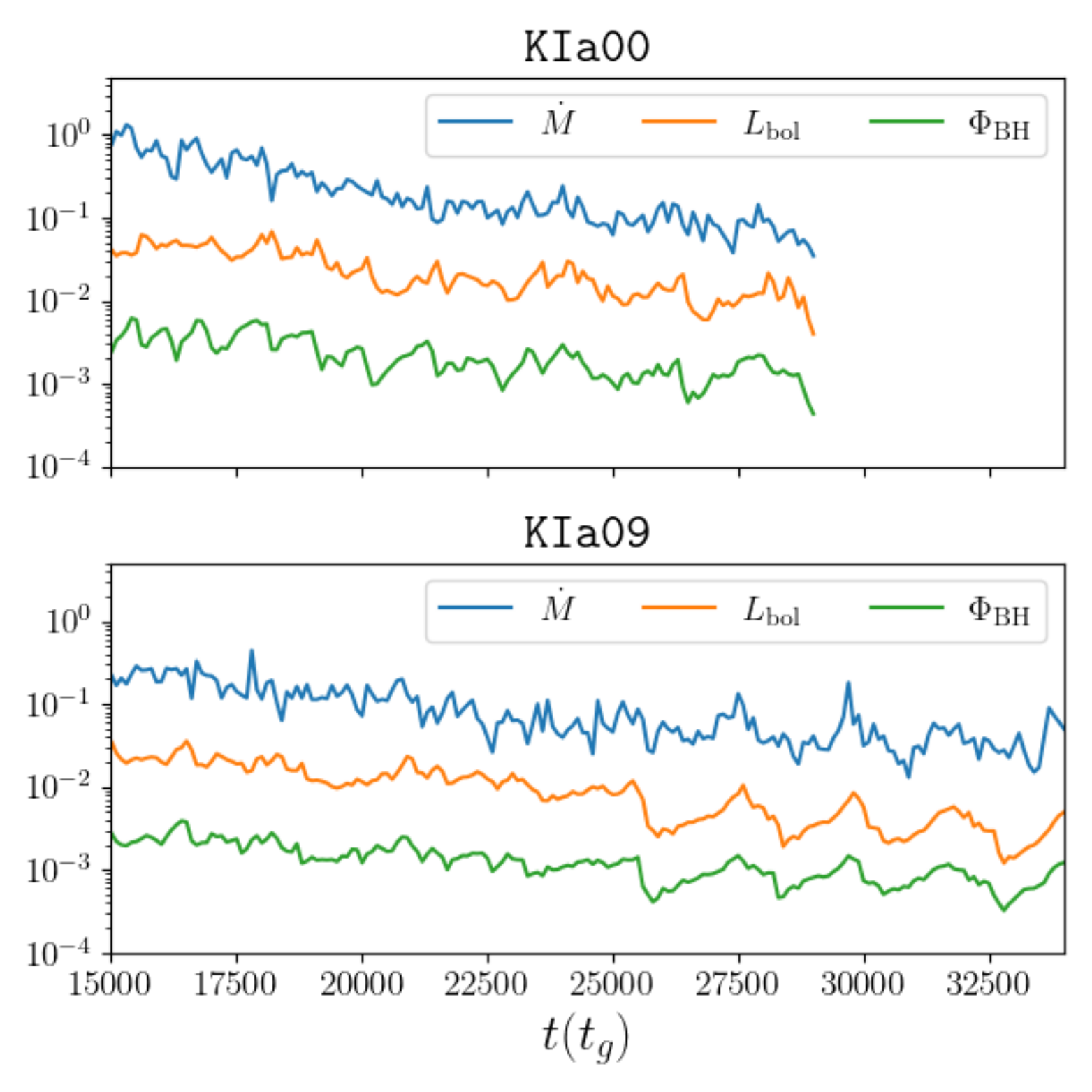}
    \caption{Here we arbitrarily scale the accretion rate $\dot{M}$, bolometric luminosity $L_{\rm{bol}}$, and magnetic flux at the BH horizon $\Phi_{\rm{BH}}$ to compare their variability in \texttt{KIa00} (top) and \texttt{KIa09} (bottom). The accretion rate, bolometric luminosity, and magnetic flux appear to show similar variability (e.g. we observe large dips at similar times); however, visually, the bolometric luminosity and magnetic flux show the strongest correlation.}
    \label{fig:variability}
\end{figure}

\begin{figure}
    \centering{}
	\includegraphics[width=\columnwidth]{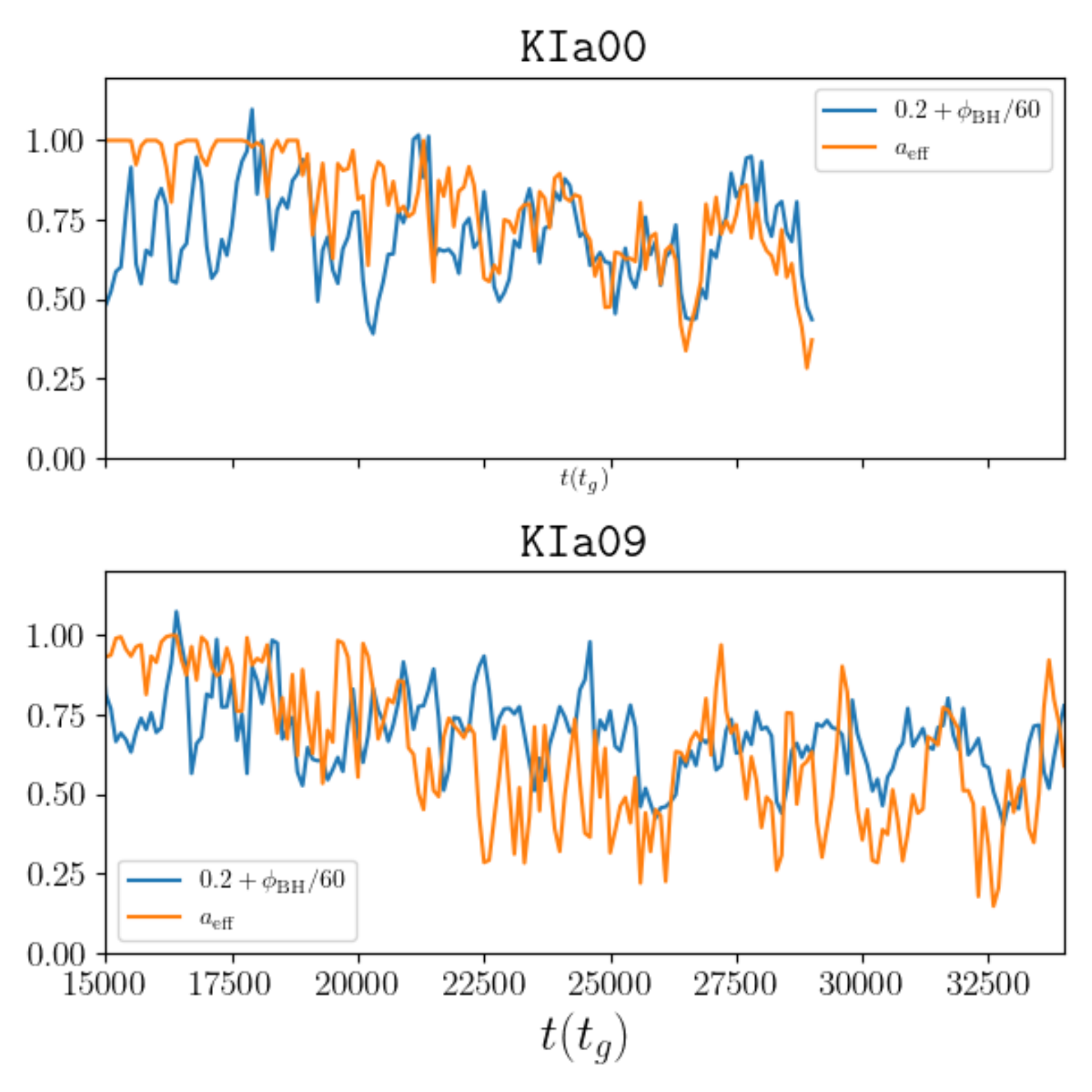}
    \caption{Here we show the fraction of the disk surface area within $r\leq 10 r_g$ that is optically thick $a_{\rm{eff}}$ over time in \texttt{KIa00} and \texttt{KIa09}. We also scale the normalized magnetic flux at the BH horizon $\phi_{\rm{BH}}$ for ease of comparison between the two parameters. In both simulations, we find that the area of the optically thick gas shows similar variability to the magnetic flux. The drop in $a_{\rm{eff}}$ appears to roughly coincide with drops in the magnetic flux due to the low density regions caused by flux eruptions.}
    \label{fig:eff_area}
\end{figure}

As shown in Figure \ref{fig:variability}, the variability of the bolometric luminosity in both \texttt{KIa00} and \texttt{KIa09} is clearly also matched by variability in the magnetic flux. We also note that \texttt{KIa09} appears to show slightly more coherent quasi-periods after $t>25,000\, t_g$. The largest dips appear in both the radiative flux and magnetic flux, and are separated by quite long times ($\delta t \approx2000\, t_g$, or nearly $10^4$\,s for our $10^6 M_\odot$ BH). The variability in the luminosity results in a decay to nearly 25\% of the peak luminosity at the minima. This suggests that MAD accretion flows may also be sources of longer period QPOs, in addition to the often cited short period QPOs associated with the spin frequency at the BH horizon (e.g. \citealt{2012JPhCS.372a2040T,2014MNRAS.437.2744T}).

As the apparent correlation between $\Phi_{\rm{BH}}$ and $L_{\rm{bol}}$ is present in both \texttt{KIa00} and \texttt{KIa09}, it is possible that the reduction in emitting area of the disk (where the disk is optically thick at the midplane) is what causes drops in the bolometric luminosity. To test this hypothesis, we compute the fraction of the disk surface area that is optically thick between $r_H$ and $10\, r_g$ as
\begin{equation} \label{eq:Aeff}
  a_{\rm{eff}}=\dfrac{\int_{r_H}^{10r_g}\int_0^{2\pi}f(\tau_\theta)r\,dr\,d\phi}{\int_{r_H}^{10r_g}\int_0^{2\pi}r\,dr\,d\phi},
\end{equation}
where
\begin{equation}
    \tau_\theta=\int_0^{\pi/2}\dfrac{\rho \kappa_{\rm{es}}}{c}u^t\sqrt{g_{\theta\theta}}\,d\theta
\end{equation}
is the angle integrated electron scattering optical depth at the midplane and $f(\tau_\theta)=1$ if $\tau_\theta>1$ and $f(\tau_\theta)=0$ otherwise. We only integrate up to $r=10\,r_g$ since the disk tends to become optically thick at larger radii, which would tend to damp oscillations in $a_{\rm{eff}}$ near the horizon if included in the integral. As we show in Figure \ref{fig:eff_area}, $a_{\rm{eff}}$ generally decreases around the same time that $\phi_{\rm{BH}}$ decreases due to magnetic islands disrupting the inner disk.

\begin{table}
    \centering
    \begin{tabular}{ c c c c}
        \hline
        \hline 
        Model & Variable 1 & Variable 2 & Correlation  \\
                & & & \\
        \hline
        \texttt{KIa00} & $L_{\rm{bol}}$ & $\Phi_{\rm{BH}}$ & $0.59\pm0.14$  \\
        & $L_{\rm{bol}}$ & $\dot{M}$ & $0.24\pm0.18$ \\
        & $\Phi_{\rm{BH}}$ & $\dot{M}$ & $0.56\pm0.22$  \\
        & $L_{\rm{bol}}$ & $a_{\rm{eff}}$ & $0.25\pm0.15$  \\
        & $\Phi_{\rm{BH}}$ & $a_{\rm{eff}}$ & $0.38\pm0.13$  \\
        & & & \\
        \texttt{KIa09} & $L_{\rm{bol}}$ & $\Phi_{\rm{BH}}$ & $0.75\pm0.13$ \\
        & $L_{\rm{bol}}$ & $\dot{M}$ & $0.32\pm0.17$ \\
        & $\Phi_{\rm{BH}}$ & $\dot{M}$ & $0.37\pm0.18$ \\
        & $L_{\rm{bol}}$ & $a_{\rm{eff}}$ & $0.25\pm0.12$  \\
        & $\Phi_{\rm{BH}}$ & $a_{\rm{eff}}$ & $0.38\pm0.12$  \\
    \hline
    \end{tabular}
    \caption{Cross-correlation coefficients (CCCs) of various quantities in which we identify significant variability.}
    \label{tab:tab3}
\end{table}

To quantify the correlation between variability in the above quantities, we compute the cross-correlation between the variables $L_{\rm{bol}}$, $\Phi_{\rm{BH}}$, $\dot{M}$, and $a_{\rm{eff}}$. For this analysis, we use data from $t=15,000 - 29,000 \, t_g$ for \texttt{KIa00} and $t=15,000 - 34,000 \, t_g$ for \texttt{KIa09}. We detrend the data using a moving average so as to leave only the short time scale variability. We then perform a stationary block bootsrap cross-correlation (e.g. see \citealt{1979AnnStat....7..1}) using $13-14$ blocks\footnote{Our chosen number of blocks is on the low end since we also had the additional requirement of preserving the auto-correlation. One would typically desire $\geq 30$ blocks, where the Central Limit Theorem is satisfied for both Gaussian and non-Gaussian sample distributions; however, we found that 13 blocks was adequate to obtain approximately Gaussian distributions of the statistic.} of equal duration in time for each model. We performed 5000 re-samples with replacement. 

We report the cross-correlation coefficient (CCC) and errors in Table \ref{tab:tab3}. We saved snapshots of the simulation every $\delta t=100\, t_g$, so we do not report lags since the sampling rate is longer than the minimum lag we expect between the mass accretion rate and luminosity, which is roughly the light crossing time $\sim30\,t_g$ for emission near the BH horizon. Although it is commonplace in statistical analysis to define a range for the CCC as strong, moderate, or weak, we do not adopt such a definition here as it is somewhat arbitrary. We find it more useful to consider the relative value of the CCC.

For \texttt{KIa00}, we find a significant correlation between $L_{\rm{bol}}$ and $\Phi_{\rm{BH}}$ with a CCC of $0.59\pm0.14$ and between $\Phi_{\rm{BH}}$ and $\dot{M}$ with a CCC of $0.56\pm0.22$. However, $L_{\rm{bol}}$ and $\dot{M}$ show a weak correlation with a CCC of $0.24\pm0.18$. For \texttt{KIa09}, the cross-correlation indeed finds that $L_{\rm{bol}}$ and $\Phi_{\rm{BH}}$ are strongly correlated with a peak CCC of $0.75\pm0.13$. Meanwhile, there is a much weaker correlation between $\dot{M}$ and $L_{\rm{bol}}$ (and $\Phi_{\rm{BH}}$).

For both \texttt{KIa00} and \texttt{KIa09}, we find that $\Phi_{\rm{BH}}$ and $a_{\rm{eff}}$ have a CCC of $0.38$; however, $L_{\rm{bol}}$ appears to be very weakly correlated with $a_{\rm{eff}}$ with a CCC of $0.25$. This suggests that the variability is not caused by thermal emission from the disk. It is possible that the variability in $L_{\rm{bol}}$ may instead be due to the flux eruptions modifying the emission properties in the jet/wind, but a more in depth investigation of the emission properties of the simulation are needed in order to test this.

Since increasing the $\sigma_M$ ceiling also increases $L_{\rm{bol}}$ in the jet region, the larger correlation between $L_{\rm{bol}}$ and $\Phi_{\rm{BH}}$ in \texttt{KIa09} when compared to \texttt{KIa00} may be due to the ceiling model. However, the fact that \texttt{KIa00}, which did not have a jet, is less effected by the ceiling on $\sigma_M$ (Figure \ref{fig:sigma_comparison}) suggests that our finding of a high CCC for $L_{\rm{bol}}$ and $\Phi_{\rm{BH}}$ in both models, when contrasted with the CCC between other variables, is a robust result.

\begin{figure}
    \centering{}
	\includegraphics[width=\columnwidth]{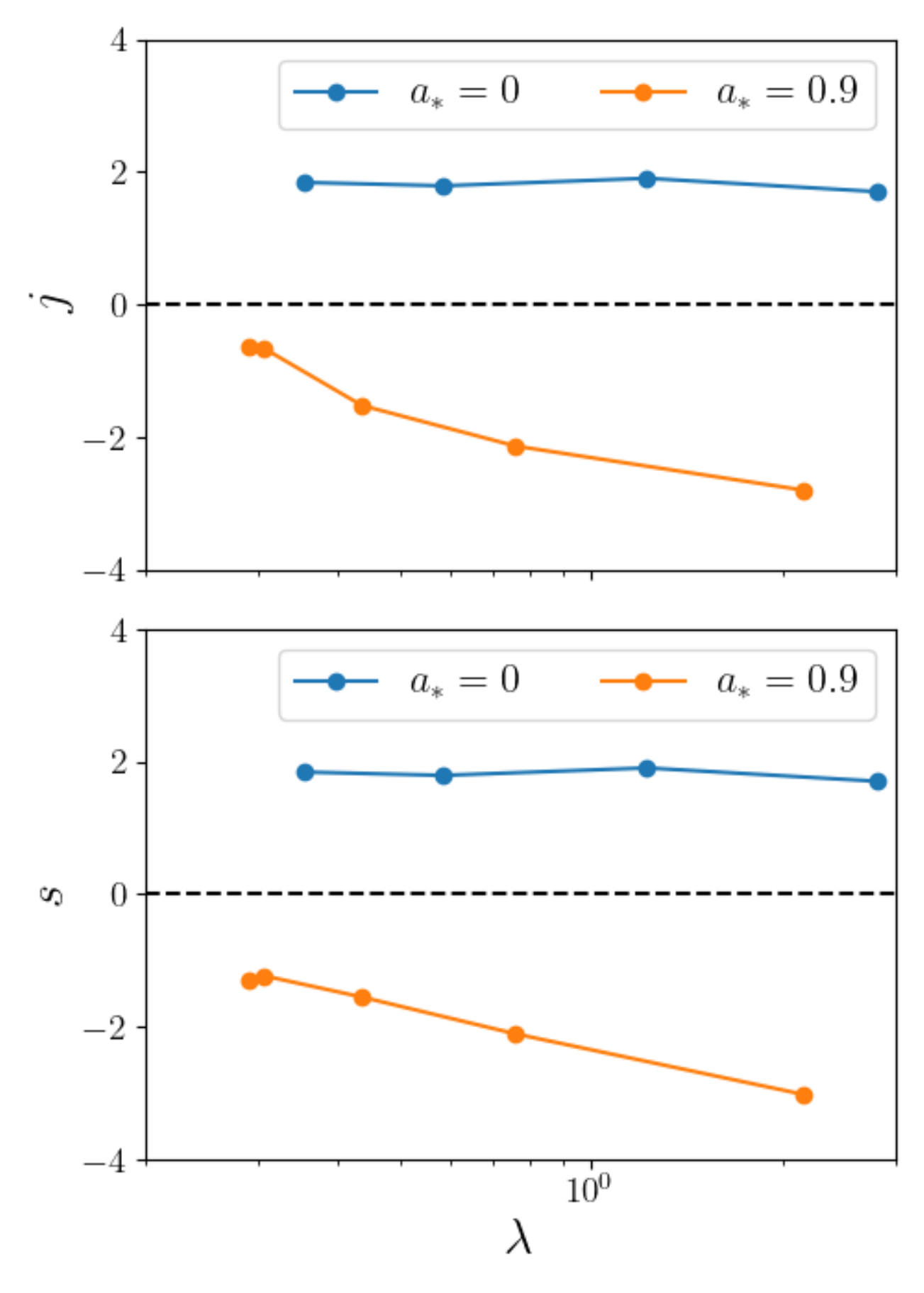}
    \caption{Here we show time-averaged specific angular momentum flux into the BH $j$ (top panel, see Equations \ref{eq:jtot}, \ref{eq:jdot}) and the dimensionless spin-up parameter $s$ (bottom panel, see Equation \ref{eq:spinup}) as a function of the accretion rate $\lambda$ for spin $a_*=0$ (\texttt{KIa00}) and $a_*=0.9$ (\texttt{KIa09}).}
    \label{fig:spindown}
\end{figure}

\begin{table}
    \centering
    \begin{tabular}{ c c c c c }
        \hline
        \hline 
        Model & Time & $\lambda$ & $j$ & $s$ \\
               & $(10^3 \, t_g)$ & & & \\
        \hline
        \texttt{KIa00} & 15-18 & 2.8 & 1.70 & 1.70\\
                       & 18-21 & 1.2 & 1.91 & 1.91\\
                       & 22-25 & 0.58 & 1.79 & 1.79 \\
                       & 26-29 & 0.35 & 1.84 & 1.84 \\ 
        \hline
        NT $(a_*=0)$   & - & - & 3.46 & 3.46 \\ 
        \hline
         & & & \\
        \texttt{KIa09} & 14-17 & 2.2 & -2.80 & -3.02\\
                       & 20-23 & 0.76 & -2.14 & -2.11 \\
                       & 23-26 & 0.44 & -1.52 & -1.55 \\
                       & 28-31 & 0.31 & -0.66 & -1.23 \\ 
                       & 31-34 & 0.29 & -0.64 & -1.29 \\ 
        \hline
        NT $(a_*=0.9)$   & - & - & 2.10 & 0.58 \\ 
    \hline
    \end{tabular}
    \caption{Tabulated values of the specific angular momentum flux into the BH ($j$), and spin-up parameter ($s$) for each model as a function of time and accretion rate ($\lambda$).}
    \label{tab:tab4}
\end{table}

\section{Spin Evolution} \label{sec:spindown}

We show the specific angular momentum $j$ of the accreted gas and the dimensionless BH spin-up parameter $s$ of each model in Figure \ref{fig:spindown}. We also present tabulated values of $j$ and $s$ as a function of time and $\lambda$ in Table \ref{tab:tab4}. For \texttt{KIa00}, we find that both $j$ and $s$ are positive, meaning that angular momentum and net spin are being gained by the BH. Furthermore, both quantities are nearly independent of the accretion rate. In the case of \texttt{KIa09}, both $j$ and $s$ have the opposite sign, indicating that angular momentum is being {\it lost by the BH}. Thus, rapidly spinning BHs, even with thin disks, spin down if accretion occurs in the MAD state. For comparison, the traditional NT disk spins up the BH for all $a_*$ values up to $0.998$ \citep{1974ApJ...191..507T}. Note that $s$ is correlated with $\lambda$ in \texttt{KIa09} and becomes less negative with decreasing $\lambda$. This suggests that, if $\lambda$ is sufficiently small, the BH may switch from spin-down to spin-up.

\begin{figure}
    \centering{}
	\includegraphics[width=\columnwidth]{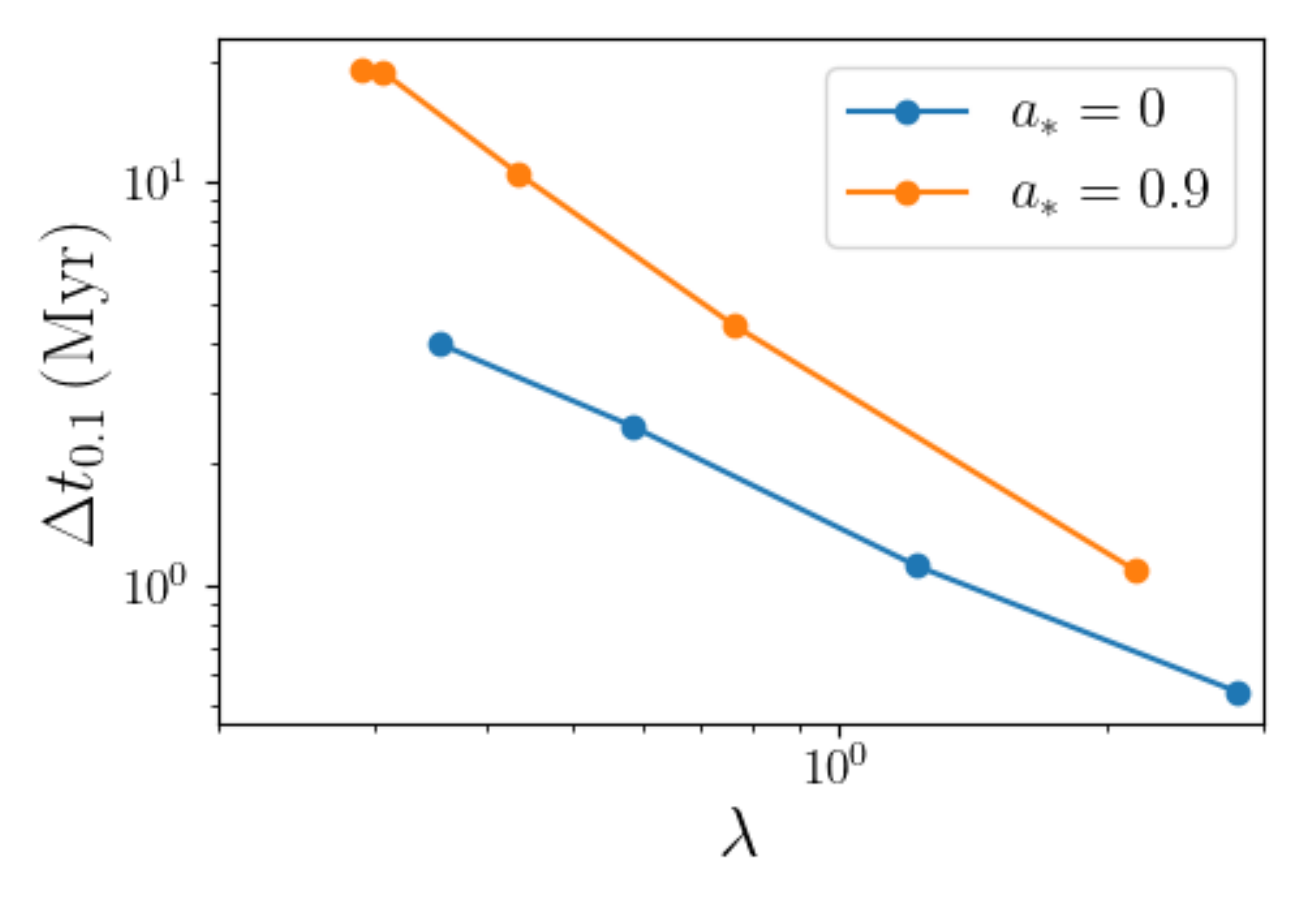}
    \caption{Here we estimate the time for the spin $a_*$ of the BH to increase/decrease by $0.1$ as a function of accretion rate $\lambda$ for spin $a_*=0$ (\texttt{KIa00}) and $a_*=0.9$ (\texttt{KIa09}).}
    \label{fig:spindowntime}
\end{figure}

An interesting question to consider is the time evolution of the BH spin. We follow the analysis in \citet{2022MNRAS.511.3795N}, which is based on \citet{2005ApJ...620...59S}, and calculate the spin evolution via:
\begin{equation}
    \dfrac{\Delta a_*}{\Delta t} = \lambda\dfrac{\dot{M}_{\rm{Edd}}}{M}s(\lambda).
\end{equation}
To calculate the change in spin over a time interval $\Delta t$, we assume $\lambda$ and $s$ are constant. 
Figure \ref{fig:spindowntime} shows the time $\Delta t_{0.1}$ for the spin to change by $|\Delta a_*| = 0.1$ as a function of $\lambda$ for the two models. The time required decreases as the accretion rate increases, as expected. For \texttt{KIa00}, this change is approximately inversely proportional to $\lambda$ since $s(\lambda)$ is nearly constant. For \texttt{KIa09}, the dependence is steeper; $\Delta t_{0.1}$ is nearly the same as for \texttt{KIa00} for large $\lambda$, but 3 times longer for $\lambda\sim0.3$ because $s$ decreases by a factor of 3. For $\lambda > 1$, the time for the BH to spin up/down for both models is roughly 1 Myr. When $\lambda\sim0.3$, $\Delta t_{0.1}$ is $\approx 4$ Myr for \texttt{KIa00} and $\approx 15$ Myr for \texttt{KIa09}. These times are much longer than the duration of a TDE, and hence not relevant for these events. The results are, however, of great interest for AGN evolution in general, and we turn next to this topic.

Since \texttt{KIa00} has $s>0$ and \texttt{KIa09} has $s<0$, there must be a spin value $a_{*,\rm eq}$ in between $a_*=0$ and 0.9 where a thin MAD disk will reach spin equilibrium with neither spin-up nor spin-down. We obtain a crude estimate of the equilibrium spin $a_{*,{\rm{eq}}}$ using the following approximations: (i) we assume that, for a given $\lambda$, $j$ varies with BH spin as $j = c_1 + c_2a_*^2$, where $c_1$ and $c_2$ are constants; this approximation is motivated by the fact that angular momentum loss via the BZ mechanism should scale as $a_*^2$, given that the BZ power varies as $\omega_H^2 \sim a_*^2$ (see Eq~\ref{eq:PBZ}), (ii) we assume that $\eta_{\rm{tot}} = c_3 + c_4 a_*^2$ for similar reasons.

We consider three accretion rates ranging from super-Eddington to sub-Eddington: $\lambda=1.2,0.58,0.35$. For these values of $\lambda$, we have estimates of $j$ and $\eta_{\rm tot}$ directly from Table~\ref{tab:tab2} for \texttt{KIa00}. In the case of \texttt{KIa09}, we interpolate $j$ and $\eta_{\rm{tot}}$ from the tabulated values. We then compute where $s(\lambda,a_*)= 0$ using fits to $j(\lambda,a_*)$ and $\eta_{\rm{tot}}(\lambda,a_*)$ vs $a_*$, using the assumed $a_*$ dependence described in the previous paragraph.
We thereby estimate that the equilibrium spin lies roughly in the range $0.5 < a_{*,{\rm{eq}}} < 0.6$ for $\lambda=0.35-1.2$. The fact that the spin-up parameter of \citet{2016MNRAS.462..636A}, whose model was also thin and MAD with $a_*=0.5$, was only $s\approx-0.4$ suggests that our estimate is not too far from what a full parameter space study may find. Our estimated equilibrium spin is approximately half the spin value expected with a traditional NT thin disk model, which is expected to drive a BH to an equilibrium spin of $a_{*,{\rm{eq}}}=0.998$ \citep{1974ApJ...191..507T}.



Eddington-level or mildly sub-Eddington accretion during a SMBHs accretion history is often cited to explain the occurrence of very massive SMBHs at high redshifts $z\sim6$ \citep{2004ApJ...613...36H,2004ApJ...614L..25Y,2005ApJ...620...59S,2005ApJ...633..624V,2006ApJ...650..669V,2007ApJ...665..107P,2009ApJ...696.1798T,2014ApJ...784L..38M,2015MNRAS.452.1922P,2016MNRAS.456.2993L}. Our models suggest that MAD accretion disks, even for thin disks, may play an important role in the spin evolution of these BHs. The time that we estimate for a BH to spin up or down is not substantial even for the sub-Eddington stages. For example, even assuming a modest duty cycle $\mathcal{D}\approx0.01$ for the BH's accretion history and Eddington-scaled accretion rates $\lambda\approx0.1-1$, the time for the BH to spin up or down to its equilibrium spin is only of the order of a few Gyr (ignoring other contributions to spin evolution such as mergers).

\section{Discussion} \label{sec:discussion}

\subsection{Comparison with other thin disk simulations} \label{sec:comparison}

In recent years, studies of radiative GRMHD simulations have begun to explore $\dot{M}/\dot{M}_{\rm{Edd}} <1$ accretion disks. In this regime, it was initially unclear if thermally stable solutions with moderate vertical extent existed, but several studies have shown that stability is indeed possible at such accretion rates \citep{2016MNRAS.459.4397S,2016MNRAS.462..636A,2018MNRAS.480.3547M,2019ApJ...884L..37L,2020MNRAS.492.1855M,2022arXiv220103526L}. \citet{2016MNRAS.459.4397S} was the first to demonstrate that stable disks can occur at $\lambda \approx 0.8$ if a strong magnetic field threads the disk. This initial solution was turbulent with a density scale height of $0.15$ while the photosphere height was substantially larger due to magnetic pressure support. Accretion occurred both through the much more dense disk (within a density scale height) and the optically thick, but less dense gas inside the photosphere.

\citet{2019ApJ...884L..37L} extended the range of simulations of magnetically supported disks to $\lambda \approx 0.6$. They again found that a stable, turbulent disk formed with accretion occuring both through the denser disk at the midplane and above/below the disk in magnetically supported gas. The density scale height was $\approx0.1$ while the photosphere height was $H\sim r$. They named this mode of accretion a `Puffy Disk'. Both \citet{2016MNRAS.459.4397S} and \citet{2019ApJ...884L..37L} found that the magnetic pressure was nearly 50-60\% of the total pressure and radiation pressure was far greater than the gas pressure. Additionally, the radiative efficiency of these solutions is similar to that of a \citet{1973blho.conf..343N} thin disk solution. We note that both authors employed non-rotating BHs with SANE accretion disks.

\begin{figure}
    \centering{}
	\includegraphics[width=\columnwidth]{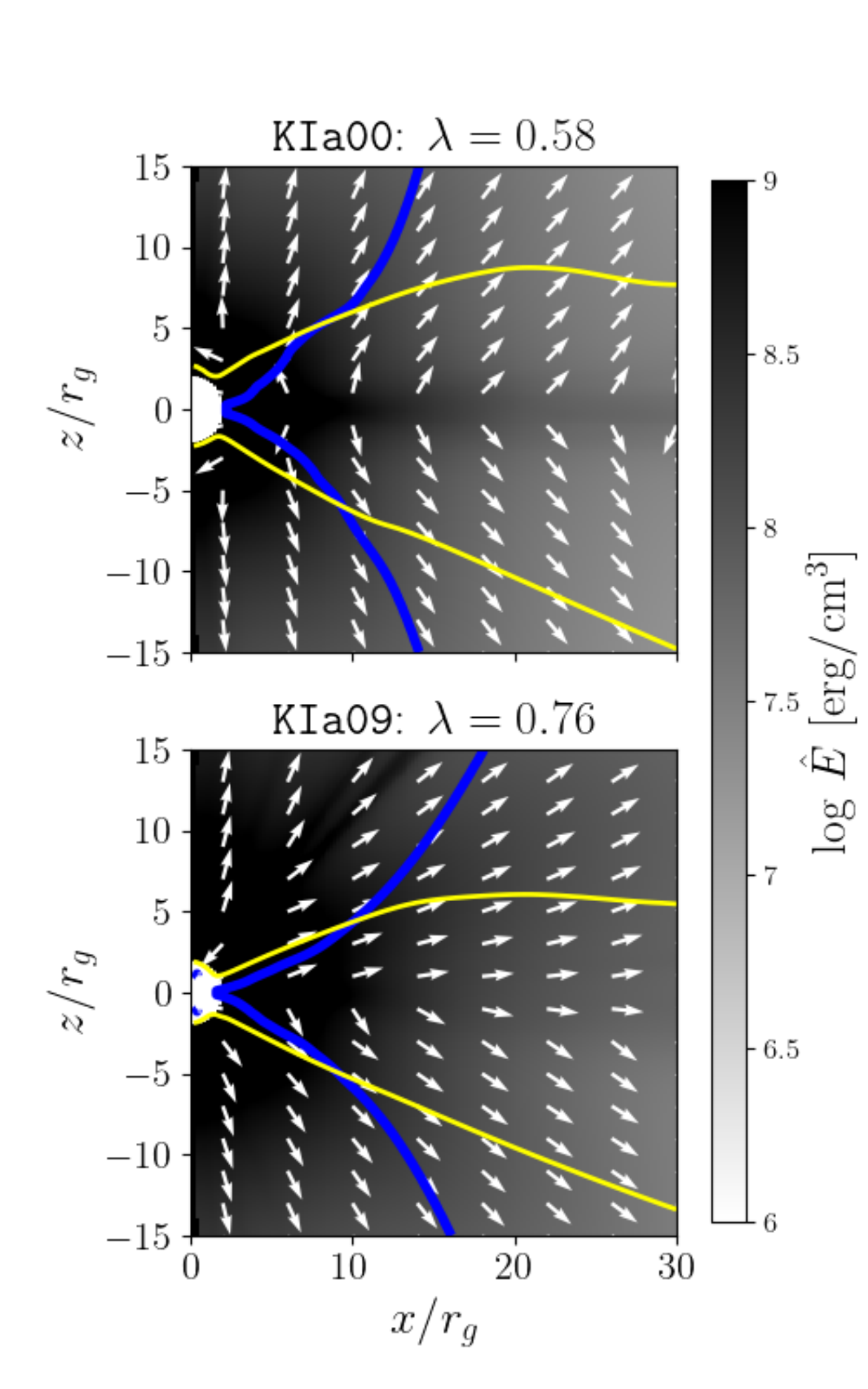}
    \caption{Here we show time-$\phi$ averaged radiation energy density (colors) and radiative flux (arrows) for both \texttt{KIa00} (top) and \texttt{KIa09} (bottom) at Eddington ratios similar to the `Puffy Disk' models presented in \citet{2016MNRAS.459.4397S,2019ApJ...884L..37L}. The electron scattering photosphere (yellow line), and magnetized jet boundary ($\sigma_M=1$, blue line) are indicated as well.}
    \label{fig:radflux}
\end{figure}

In comparing our $a_*=0$ model \texttt{KIa00} with the models presented in \citet{2016MNRAS.459.4397S,2019ApJ...884L..37L}, we find that our thin MAD solution is significantly more efficient, with $\eta_{\rm{rad}}\approx 11.6-12.2\%$ for sub-Eddington phases of the evolution. The density scale height of \texttt{KIa00} during the sub-Eddington stages ($H/R\approx 0.15-0.17$) is also slightly larger in general. We also find a considerable difference in the behaviour of the radiative flux (see Figure \ref{fig:radflux}). Both \texttt{KIa00} and \texttt{KIa09} exhibit radiative flux which escapes nearly vertically from the mid-plane even below the electron scattering photosphere. This is the opposite of the behaviour in \citet{2016MNRAS.459.4397S,2019ApJ...884L..37L}, who both found that the radiative flux was advected towards the BH beneath the photosphere. This difference is possibly due to the different angular momentum transport mechanisms, as accretion in MAD disks is driven by magnetic Rayleigh-Taylor instabilities (but see \citealt{2022MNRAS.511.2040B} who argue instead that radial convective/interchange instabilities dictate the flow structure in MADs) whereas SANE models such as the `Puffy Disk' model accrete through MRI.

It is interesting that our radiative efficiency is higher, but there are important differences between our models and the `Puffy Disk'. For instance, they found that additional radiation from heightened viscous heating inside the ISCO cannot escape due to radial advection inward, which does not appear to be the case in our models. Also, the photosphere scale height in our models is significantly smaller than $H\sim r$, which implies that the conversion of radiative energy into kinetic energy will be less efficient than in a 'Puffy Disk', so more radiative energy will ultimately escape. The magnetized wind carries gas which holds a significant amount of radiation from the disk with it and it seems that unlike in the super-Eddington case (see \citealt{2015MNRAS.454L...6M}), this radiation escapes freely with very little absorption due to the low density of the wind. We note that for our model \texttt{KIa09} the spin is slightly higher ($a_*=0.9$ versus $a_*=0.8$) than the simulation of \citet{2015MNRAS.454L...6M}. Furthermore, the fact that we must measure the energy outflow at small radii ($r\geq 30 \, r_g$) may impact our results. For instance, \citet{2015MNRAS.454L...6M} cite an order of two decrease in their radiative efficiency when they account for the energy conversion out to large radii.

\citet{2018MNRAS.480.3547M} presented radiative GRMHD simulations of thin, MAD accretion disks around rotating ($a_*=0.5$) BHs. They simulated various disks spanning accretion rates from $\lambda \approx 0.2-0.5$ and confirmed that their disks were stable during the simulated time. They find total efficiencies of $\sim15-20$\% with the lowest efficiency corresponding to the lowest accretion rate model and vice versa, which is substantially larger than the \citet{1973blho.conf..343N} thin disk efficiency of 8.2\%. They also cite radiative efficiencies of $\sim 2.4-2.9\%$, making their models rather radiatively inefficient (that is, most of their large total efficiency is in mechanical energy outflow).

Our results for \texttt{KIa09} are similar to the findings of \citet{2018MNRAS.480.3547M} in the sense that the extraction of spin energy leads to a total efficiency that is significantly larger than that of the NT thin disk efficiency even at highly sub-Eddington rates. However, we observe a large difference with regards to the photsophere scale height and the overall energy output. \citet{2018MNRAS.480.3547M} found a photosphere height that was significantly larger than the density scale height, which suggests that radiation energy would have been efficiently converted into kinetic energy. This appears to be true of their simulations as the sum of their reported jet and wind efficiency for their model RADHR ($14.7\%$) is nearly five times larger than the radiative efficiency ($2.9\%$). It is worth noting that the disk at $r\gtrsim16\, r_g$ exhibited turbulence in all of their models since the disk was only MAD up to $\sim16 \, r_g$. The application of disk injection at $R_{\rm{inj}}=40\,r_g$ in our simulations, and the fact that our entire disk below the injection radius becomes MAD, may be a possible explanation for why our simulation behaves differently. Perhaps the extent to which the disk is MAD plays a role in the photosphere scale height and the total radiative efficiency in thin MAD disks.

A recent study of truncated disks by \citet{2022arXiv220103526L} provides a similar model to \texttt{KIa09}. Their numerical model was run with $a_*=0.9375$, $\lambda\approx0.35$, and the magnetic flux at the BH horizon saturated at $\phi_{\rm{BH}}\approx 30$, as in the case of our simulations. However, the magnetic pressure supported inner disk is somewhat thinner in their simulations with $H/R \sim 0.05-0.2$. They also evolved simulations using a two temperature electron model, which leads to cooler electrons in the jet. They confirm the possibility of large total efficiency in a thin MAD. They find $\eta_{\rm{tot}}\sim90\%$, which is not very different from the values $\eta_{\rm{tot}}\approx 70-100\%$ that we find in model \texttt{KIa09}. However, they find a smaller radiative efficiency  $\eta_{\rm{rad}}\sim20\%$ compared to our $\eta_{\rm{rad}}\sim50\%$ at $\lambda\approx0.3$. The results of \citet{2020ApJ...904..117K} suggest that the rather large difference in $\eta_{\rm{rad}}$ between our work and \citet{2022arXiv220103526L} is possibly due to the fact that they modeled a two temperature plasma, whereas our simulations assume a single temperature. However, we also demonstrated that the numerical ceiling on $\sigma_M$ impacts the radiative efficiency in \texttt{KIa09}. In addition, our choice of $\sigma_M\leq100$ is larger than the maximum values of $\sigma_M\leq 25$ and $50$ implemented by \citet{2022arXiv220103526L}. An increase in radiative efficiency with larger $\sigma_M$ was also shown in their work (see Figure 8 in \citealt{2022arXiv220103526L}). Therefore, we believe the different numerical floors and the inclusion of a two-temperature plasma can account for the factor of $\gtrsim2$ difference in our radiative efficiency when compared with \citet{2022arXiv220103526L}.

\subsection{Variability}

Simulations of MADs (e.g. \citealt{2008ApJ...677..317I}) show significant variability in the mass accretion rate.
Because of magnetic flux eruptions,
a quasi-periodic accretion rate (and thus light curve) may be expected. Although the gas can slip past the magnetic field through 3D RT instabilities and accrete, there may be a connection between variability and the radius to which the disk is MAD since the radius at which magnetic buoyancy can support the gas grows with magnetization.

Our simulations are unique in comparison to other MAD disk simulations in the sense that we inject a Keplerian disk which is forced to be MAD. Due to the boundary conditions implemented in this work, the disk can only be MAD up to $R_{\rm{inj}}$ by construction. The fact that we found structured variability in the bolometric luminosity for \texttt{KIa09} that closely matched that of the magnetic flux suggests that quasi-periods linked to flux eruptions may appear in thin MAD accretion disks.

If MAD disks can maintain large scale poloidal fields up to radii beyond the horizon, our model \texttt{KIa09} suggests quasi-periods of the order $2\times 10^3\, t_g$, accompanied by a luminosity variation with minima of $\lesssim25\%$ of the peak value, may be possible. This could have implications for well-known BH X-ray binaries (BHXBs) that show a variety of time-variable states, e.g., GRS~1915+105 \citep{2000A&A...355..271B} as well as some TDEs. 

Type-B QPOs are long frequency QPOs in BHXBs that coincide with the soft intermediate state (SIMS) and appear with a typical frequency ranging from $4-6$ Hz and fractional root mean squared (rms) amplitude of less than $10\%$ (\citealt{2010LNP...794...53B}). The SIMS and the occurrence of type-B QPOs is possibly associated with the presence of transient jets \citep{2004MNRAS.355.1105F,2005ApJ...632..504C}, which suggests our model \texttt{KIa09} may be analogous to the SIMS. Scaling the BH mass in \texttt{KIa09} to $10\,M_\odot$ suggests a QPO frequency of $\sim10$ Hz for a stellar mass BH analog, which is remarkably similar to the frequency range of type-B QPOs. We note, however, that the fractional rms amplitude in $L_{\rm{bol}}$ is $\sim50\%$ in \texttt{KIa09}, which suggests that our model may be too violently variable to explain type-B QPOs.

The fact that the variability period is $\delta t\approx 2\times10^3\, t_g$ and is similar to $T_{\rm{acc}}(R_{\rm{inj}})\approx 2500\, t_g$ suggests a connection between the period and the truncation radius of the MAD disk. However, we did not explore the effects of different choices of $R_{\rm{inj}}$ on the variability in this work. Assuming $M_{\rm{BH}}=10\, M_\odot$, a model where $\delta t \sim T_{\rm{acc}}$ would require the radial edge of the MAD disk to extend to larger radii in order to explain lower frequency QPOs.

The most well studied TDE which is thought to have gone MAD is \textit{Swift} J1644+57 \citep{2014MNRAS.437.2744T}, which showed variability on a range of timescales with both short period QPOs at $\sim 200$ s \citep{2012Sci...337..949R} and long period dips in the light curve on time scales of $\sim 10^6$ s \citep{2012MNRAS.422.1625S}. The QPO period in \texttt{KIa09} is $\approx 10^4$ s, but scaling $\delta t$ with $M_{\rm{BH}}$ and $R_{\rm{inj}}$ allows for longer periods capable of explaining the $10^6$ s variability in \textit{Swift} J1644+57. In addition, the flux dips of the long period variability had minima of $\lesssim50\%$ of the peak value, which is similar in magnitude to the dips that we find in \texttt{KIa09}.

One complication with the hypothesis that a MAD disk may drive longer period QPOs in the context of TDEs is that the magnetic flux required for the disk to be MAD beyond the horizon is quite large. This is a substantial astrophysical hurdle that makes the MAD scenario for TDEs difficult to begin with since the flux provided by the star is too small \citep{2014MNRAS.445.3919K}. Nevertheless, given the success of MAD disks in explaining events such as \textit{Swift} J1644+57, one can assume substantial flux was indeed brought in, whether from a fossil disk or from a dynamo process.

\subsection{Implications for \textit{Swift} J1644+57} \label{sec:comparison}

Our model \texttt{KIa09} is the first GRRMHD simulation of a sub-Eddington MAD disk around a SMBH. Our results demonstrate that a substantial jet power efficiency of $\eta_{\rm{jet}}\sim18\%$ can be expected  even when $\lambda \sim 0.3$, provided the disk is MAD and the BH spins rapidly. At such low accretion rates, model \texttt{KIa09} has a lower scaled magnetic flux of $\phi_{\rm BH} \sim 25-30$ compared to $\phi_{\rm BH} \sim 50$ for highly super-Eddington accretion rates \citep{2019MNRAS.483..565C}. This does cause a modest decrease in jet efficiency (see the BZ power scaling in Eq~\ref{eq:PBZ}) at the lowest sub-Eddington rates that we have explored. However, the reduction in $\eta_{\rm{jet}}$ is not sufficient to explain the apparent shut-off of the jet in \textit{Swift} J1644+57 assuming the fallback rate after $\sim500$ days was similar to the minimum mass accretion rate we probe in our simulations.

One possibility is that the disk in \textit{Swift} J1644+57 did maintain enough magnetic flux to remain MAD, but the disk became much thinner than in our simulation. The fitting formula provided by \citet{2016MNRAS.462..636A} suggests an association between $H/R$ and the extracted jet power: $\eta_{\rm{jet}}\propto (H/R)^2$. This is also suggested by the BZ power in Equation \eqref{equ:etaBZ} since various authors demonstrated through numerical simulations that $\phi_{\rm{BH}}\propto H/R$ for thick accretion disks \citep{2010MNRAS.408..752P,2011MNRAS.418L..79T,2012JPhCS.372a2040T}.

Based on the disk height evolution in \texttt{KIa09} (see the bottom panel in Figure \ref{fig:KIa09scalars}), the above scenario would only be realized at even lower accretion rates than we have considered. In this context, the accretion rate at the time that the X-ray flux shut off in \textit{Swift} J1644+57 may have been significantly smaller than the cutoff mass accretion rate of $\lambda\sim0.3$ assumed by \citet{2014MNRAS.437.2744T}.

For instance, if we consider $H/R\sim 0.03$, which is nearly an order of magnitude smaller than the disk height at $\lambda=0.29$ in \texttt{KIa09}, the jet efficiency assuming $\eta_{\rm{jet}}\propto (H/R)^2$ is only $\sim2\%$. The trend that we find in $H/R$ with decreasing $\lambda$ suggests $(H/R)\propto \lambda^{0.5}$ for $\lambda=0.29-0.44$. Assuming this behavior holds for yet smaller accretion rates, the jet power may decrease to $\eta_{\rm{jet}}\lesssim2\%$ if $\lambda \lesssim 0.02$. A potential difficulty with this scenario is that, while a lower accretion rate and correspondingly smaller $H/R$ predicts a steep reduction in the jet power, we expect the mass accretion rate itself to decline smoothly with time in a TDE. The tidal stream is assumed to feed mass into an accretion disk at a rate $\dot{M}_{\rm{fb}}\propto (t/t_{\rm{fb}})^{-5/3}$, where $\dot{M}_{\rm{fb}}$ is the fallback rate and $t_{\rm{fb}}$ is the fallback time. Assuming $\lambda \propto (t/t_{\rm{fb}})^{-5/3}$, it would take nearly $4t_{\rm{fb}}\approx 124$ days for $\lambda$ to decrease from $0.3$ to $0.02$ if $M_{\rm{BH}}=10^6\,M_\odot$. It is hard to see how this can lead to as abrupt a decline in the X-ray flux as was seen in \textit{Swift} J1644+57 since the X-ray flux declined by a factor of 15 in only 25 days \citep{2012ATel.4398....1S}, which is substantially shorter than $4t_{\rm{fb}}\approx 124$ days.

We caution that the $\eta_{\rm{jet}}$ fits in \citet{2016MNRAS.462..636A} that we used in the above calculation were only provided over data sampling $H/R\gtrsim 0.05$ so we have assumed that $\eta_{\rm{jet}}\propto (H/R)^2$ at lower disk scale heights. Also of note is the fact that \citet{2022arXiv220103526L} report $\eta_{\rm{jet}}\sim50\%$ for a disk with $H/R\sim 0.05$, which suggests that the correlation of $\eta_{\rm{jet}}$ with $H/R$ may not hold when MAD disks are simulated with extreme resolution. \citet{2022arXiv220103526L} implement a grid with substantially higher resolution in the disk midplane (more than a factor of 10 in $\theta$) compared to simulations presented in both \citet{2016MNRAS.462..636A} and this work.

The other possibility is that the magnetic flux diffuses away from the BH horizon to larger radii once $\lambda$ falls below some threshold value. That is, the MAD state is no longer permitted. Such diffusion has not been seen in any simulation so far. However, it is possible that diffusion occurs on a longer time scale than the simulations can explore. For instance, \texttt{KIa09} was run for a mere $\sim2$ days (and spent $\sim 0.3$ days at $\lambda \lesssim0.3$) while the X-ray flux in \textit{Swift} J1644+57 declined by a factor of 15 in $\sim25$ days \citep{2012ATel.4398....1S}. In this scenario, the disk would switch from MAD to SANE below a critical $\lambda$ and the jet and its associated X-ray emission would be lost. Considering that all stable sub-Eddington $\lambda \lesssim 1$ disk simulations so far have included strong magnetic fields (even the `Puffy Disk' model which is technically not MAD), one wonders whether the loss of magnetic field might cause the disk to become thermally unstable. If it does, the disk will likely collapse quickly to a very small scale height since magnetic pressure support would be lost. Since the time scale for the disk to collapse due to becoming thermally unstable is short, this scenario is currently the most favorable for explaining the sudden decline in the X-ray in \textit{Swift} J1644+57.

In summation, we believe that diffusion of the magnetic flux away from the horizon and the disk no longer being MAD must coincide with the disk becoming modestly sub-Eddington to explain the X-ray shut off in \textit{Swift} J1644+57 (and possibly \textit{Swift} J2058+05). It is conceivable that the required magnetic flux  diffusion would become evident with longer simulations, and we plan to consider this in a future study.

\section{Conclusions} \label{sec:conclusions}

We have simulated the evolution of a small scale, thin, MAD accretion disk around both non-spinning (model \texttt{KIa00}) and spinning (model \texttt{KIa09}) SMBHs and tracked the evolution from mildly super-Eddington to sub-Eddington mass accretion rates. Our conclusions are as follows: 

\begin{enumerate}
    \item Throughout the evolution, the disk is stable against thermal collapse because of the large magnetic pressure in the disk.
    
    \item Both of our models are highly radiatively efficient during the sub-Eddington phase. The spin $a_*=0$ model \texttt{KIa00} reaches $\eta_{\rm{rad}}\approx 11.6-12.2\%$ and the spin $a_*=0.9$ model \texttt{KIa09} reaches $\eta_{\rm{rad}}\approx 45.4-50.2\%$. These are significantly larger than the NT thin disk efficincies $\eta_{\rm{rad}}=5.72\%$ and $\eta_{\rm{rad}}=15.58\%$, respectively, for the two spins. The total efficiency also shows significant deviations from NT during the sub-Eddington phase, with $\eta_{\rm{tot}}\approx 15.1-15.9\%$ for \texttt{KIa00} and $\eta_{\rm{tot}}\approx 64.3-68.3\%$ for \texttt{KIa09}.
    
    \item The jet efficiency in \texttt{KIa09} decreases from $\eta_{\rm{jet}}\approx38.7\%$ to $\eta_{\rm{jet}}\approx17.7-18.4\%$ as the disk becomes sub-Eddington, meaning that, even at low accretion rates, there is substantial jet power in a mildly relativistic jet driven by the extraction of spin energy. We only measure the lorentz fact of the jet at $r=30\,r_g$ due to the our application of the mass injection scheme used to vary the mass accretion rate over time. However, the asymptotic lorentz factor suggests that the outflow may reach highly relativistic speeds of $\gamma_\infty\gtrsim30$ at $r\gg r_g$ across all $\lambda$. This jet is powered by the relatively strong magnetic flux $\phi_{\rm{BH}}\approx30$ in spite of the decreasing accretion rate and disk scale height.
    
    \item For \texttt{KIa09}, we find a best fit jet profile power-law $k\approx0.43-0.53$, which is approximately parabolic, within $r<100\, r_g$ when $\lambda < 1$. This is consistent with measurements of high state AGN jets on the sub-parsec scale.
    
    \item Both \texttt{KIa00} and \texttt{KIa09} are highly variable. We find significant correlation between the escaping luminosity and the magnetic flux in both models, which suggests that the flux eruptions characteristic of the MAD state may be an important source of QPOs. We found that the fraction of the disk surface area that is optically thick within $10\,r_g$, $a_{\rm{eff}}$, is very weakly correlated with $L_{\rm{bol}}$ which suggests the cause for the correlation between $L_{\rm{bol}}$ and $\Phi_{\rm{BH}}$ is not blackbody disk emission.
    
    \item \texttt{KIa09} exhibits coherent quasi-periodic variability with a period of $\sim 2000\, t_g$ in the final $\sim 10,000\, t_g$ of the simulation. The luminosity decreases to $\lesssim 25\%$ of its peak during each dip. Such variability may be useful in interpreting long period variability in BHXBs and jetted TDEs such as \textit{Swift} J1644+57.
    
    \item For all mass accretion rates $\lambda$ that we have studied, the BH in \texttt{KIa00} ($a_*=0$) spins up with a positive spin-up parameter $s\sim2$. In contrast, the BH in \texttt{KIa09} ($a_*=0.9$) spins down with $-3\lesssim s\lesssim -1$. The spin-down rate becomes larger with increasing $\lambda$.
    
    \item For accretion disks in the MAD state with mass accretion rates $\lambda=0.35-1.2$, we estimate that the BHs will, with continued accretion, reach an equilibrium spin in the range $0.5\lesssim a_{*,{\rm{eq}}}\lesssim0.6$. This value of $a_*$ is only about half the equilibrium spin, $a_{*,{\rm{eq}}}=0.998$, expected with the standard NT thin disk model.
\end{enumerate}

\section*{Acknowledgements}

We are grateful to Koushik Chatterjee, Josh Speagle, and Matthew Liska for useful comments and suggestions. We also thank the anonymous reviewer for many useful suggestions which improved the quality of this work. This work was supported in part by NSF grant AST-1816420, and made use of computational support from NSF via XSEDE resources (grant TG-AST080026N). This work was carried out at the Black Hole Initiative at Harvard University, which is supported by grants from the John Templeton Foundation and the Gordon and Betty Moore Foundation.

\section*{Data Availability}
The data underlying this article will be shared on reasonable request to the corresponding author.



\bibliographystyle{mnras}
\bibliography{main}




\appendix

\section{Effects of Boundary Condition Choice Behind Injection Region} \label{sec:appA}

In this section, we quantify whether or not the choice of boundary conditions has a significant impact on the solution of the simulation in the regions where we analyze the output. We perform a rerun of \texttt{KIa09} in 3D from $t=30,000-34,000\, t_g$ with a reflecting boundary condition on $\bold{B}$ in the 2 radially exterior ghost cells of the injection region ($\bold{B}_{\rm{GC}}$ henceforth). We then compare diagnostics with the same time period in \texttt{KIa09} with $\bold{B}_{\rm{GC}}=0$ to evaluate whether or not key results presented in this work would be significantly impacted by this choice.

\begin{figure}
    \centering{}
	\includegraphics[width=\columnwidth]{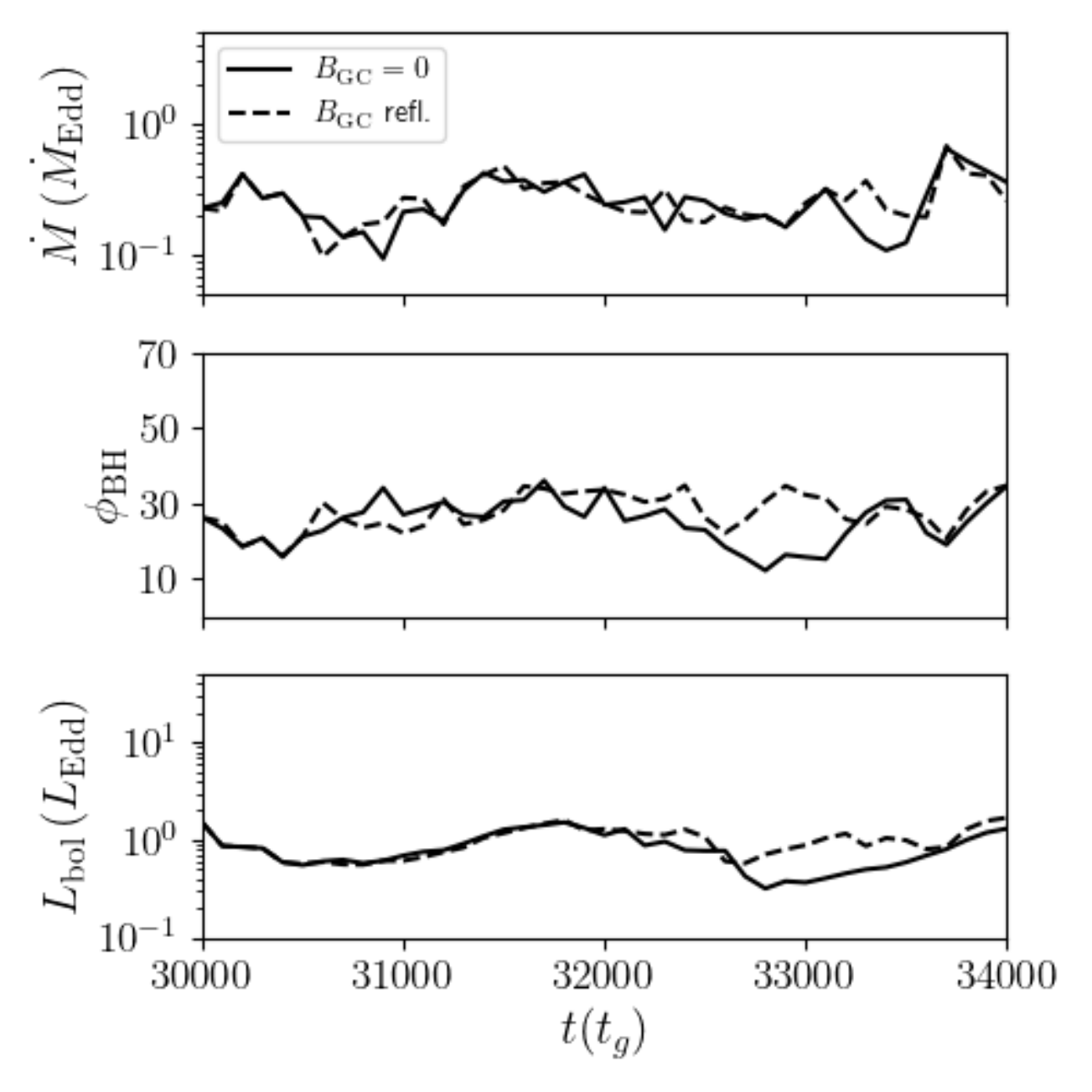}
    \caption{Here we compare $\dot{M}$, $\phi_{\rm{BH}}$ and $L_{\rm{bol}}$ in \texttt{KIa09} for boundary condition choices $\bold{B}_{\rm{GC}}=0$ (solid line) and $\bold{B}_{\rm{GC}}$ reflected (dashed line). As described in the text, we find that the simulations have similar behaviour, but there are some deviations due to the turbulent nature of the system.}
    \label{fig:BC_comparison}
\end{figure}

In Figure \ref{fig:BC_comparison}, we show the evolution of $\dot{M}$, $\phi_{\rm{BH}}$, and $L_{\rm{bol}}$ for both boundary condition choices using $\phi$-averaged data. As expected, the evolution is similar although there are minor deviations in the rerun as the system is turbulent. The most significant deviation is in $\phi_{\rm{BH}}$ and $L_{\rm{bol}}$ around around $33,000\, t_g$ due to a more significant flux eruption in the $\bold{B}_{\rm{GC}}=0$ simulation data. This is not unusual as the system is turbulent and reruns of the same state will yield slightly different behavior on short time scales.

\begin{figure}
    \centering{}
	\includegraphics[width=0.5\columnwidth]{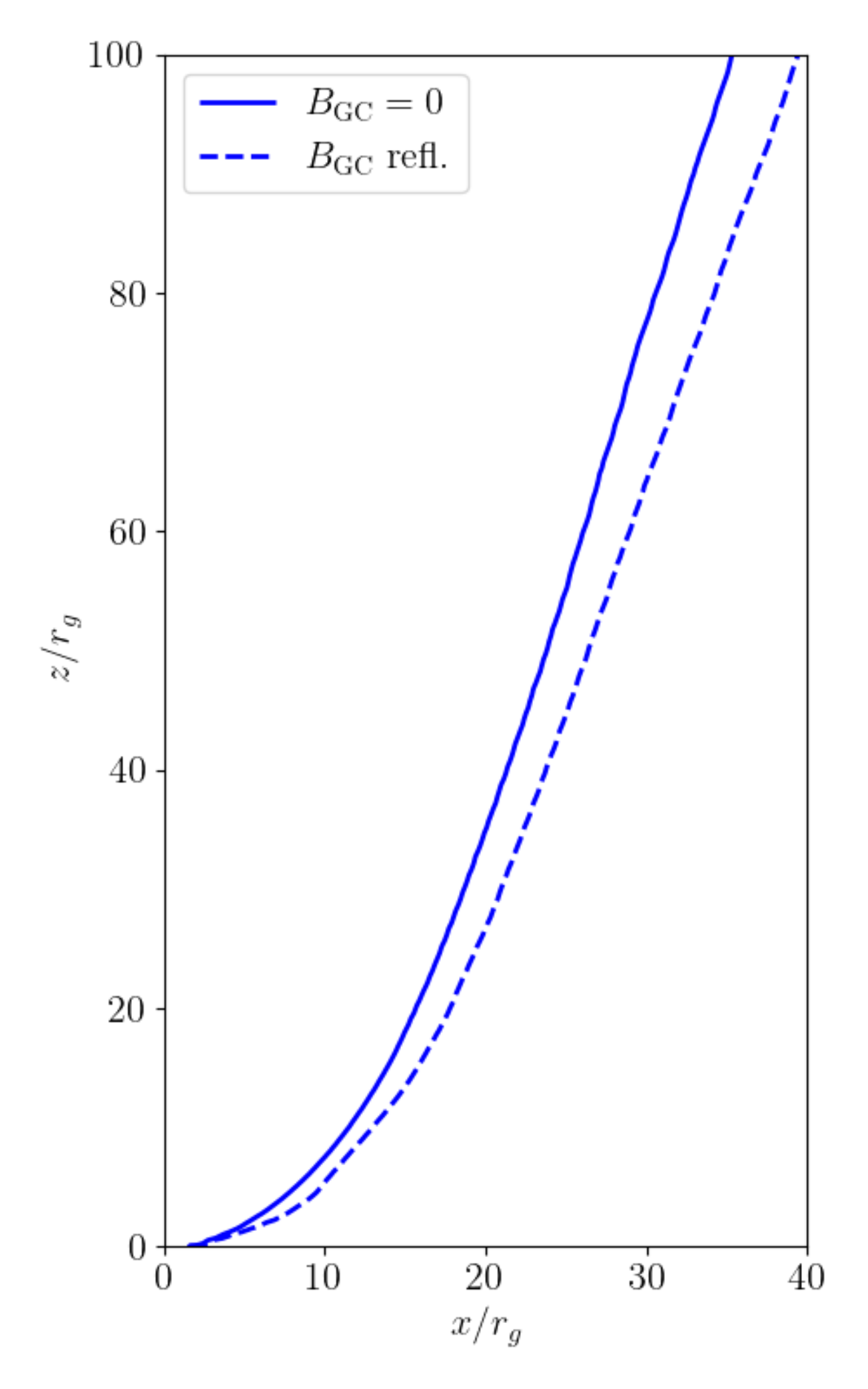}
    \caption{Here we compare the $\sigma_M=1$ contour for the top jet in \texttt{KIa09} for boundary condition choices $\bold{B}_{\rm{GC}}=0$ (solid line) and $\bold{B}_{\rm{GC}}$ reflected (dashed line). Note that we have time-averaged over $t=30,000-34,000\, t_g$ and $\phi$-averaged over $2\pi$. We find that the jet shape is roughly the same, although the $\bold{B}_{\rm{GC}}=0$ jet is slightly less wide in a time averaged sense due to the more extreme flux eruption event around $t=33,000\,t_g$.}
    \label{fig:BC_heatmap}
\end{figure}

We also compare the $\sigma_M=1$ profile of the jet using data averaged in $\phi$ and time averaged over $t=30,000-34,000\, t_g$ (Figure \ref{fig:BC_heatmap}). The overall jet profile is similar, though we note that the sharp drop in $\phi_{\rm{BH}}$ around $33,000\,t_g$ leads to a slightly less wide jet in a time-averaged sense due to the jet narrowing during low power states similar to Figure \ref{fig:KIa09jet}. With that minor deviation aside, the jet behaves in a similar manner during the rerun. Indeed, we find that a measure of the jet profile for the top (bottom) jet over the time period $t=30,000-34,000\, t_g$ yields $k\approx0.45$ (0.47), which is in the range of values shown in Figure \ref{fig:jetedge} for the sub-Eddington phase.

Lastly, we measured the disk scale height at $r=5\, r_g$ and find $(H/R)_{r=5r_g}\approx 0.12$, which is similar to what we find over the same time period in Table \ref{tab:tab2}.

As expected, the effect of our boundary condition choice for $\bold{B}_{\rm{GC}}$ on the evolution of the disk and jet appears to be minimal. As long as a large flux of poloidal magnetic field is maintained (and a MAD state is maintained), the behaviour is generally the same. As noted in the text, this choice of boundary conditions $\bold{B}_{\rm{GC}}=0$ was made to avoid the injection of magnetic field of the opposite sign through the corner ghost cells.

\section{Comparison of $\sigma_M$ in Models} \label{sec:appB}

\begin{figure}
    \centering{}
	\includegraphics[width=\columnwidth]{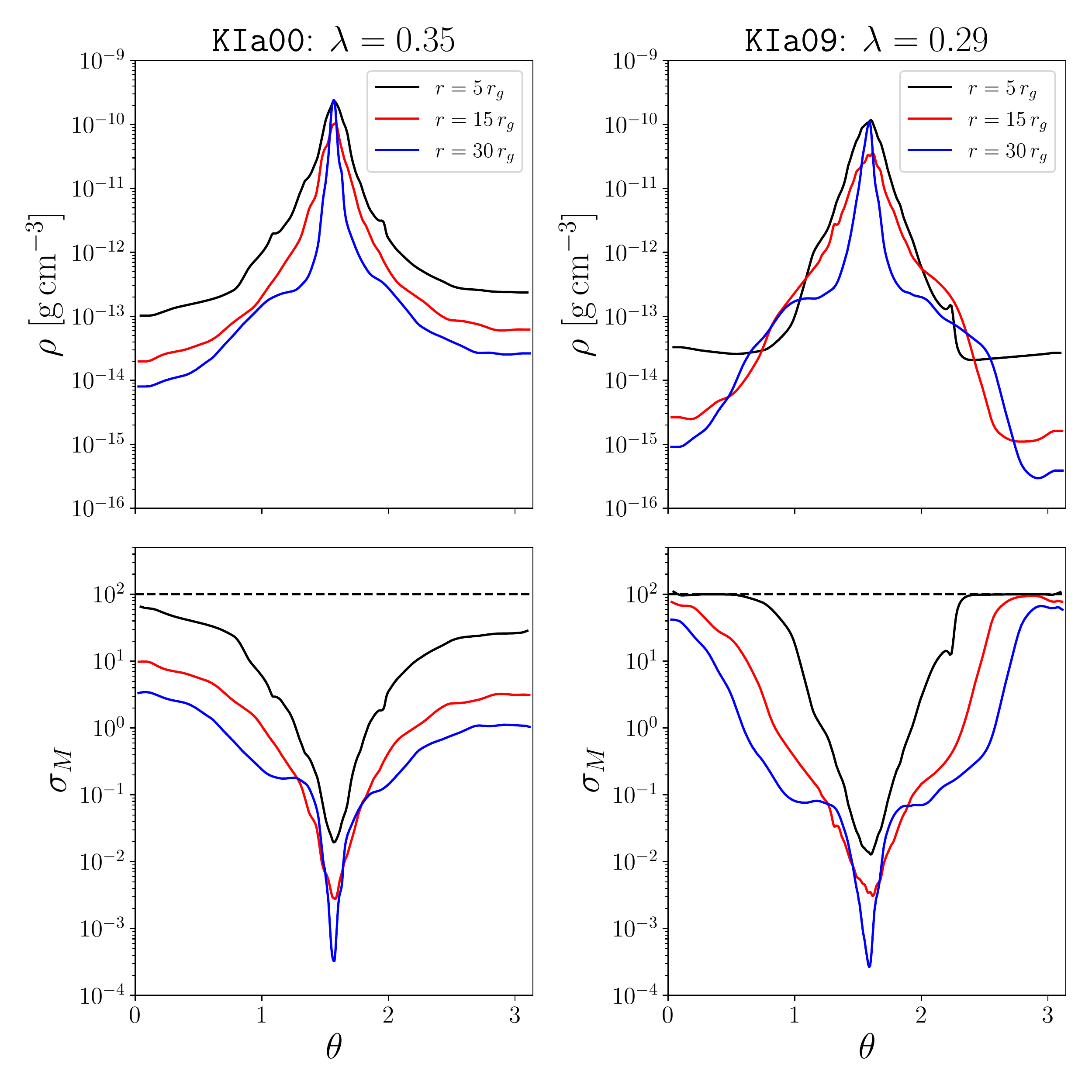}
    \caption{We show angular profiles of the gas density ($\rho$, top) and magnetization ($\sigma_M$, bottom) for model \texttt{KIa00} (left) and \texttt{KIa09} (right). We show the maximum $\sigma_M=100$ as a dashed horizontal line. In \texttt{KIa09}, the implementation of the numerical ceiling $\sigma_M\leq100$, especially at $r=5\,r_g$, leads to artificially high density gas in the funnel and a flat density profile in $\theta$. \texttt{KIa00} generally does not artificially add mass in the jet since $\sigma_M$ is below 100 throughout most of the funnel even at radii as small as $r=5\, r_g$.}
    \label{fig:sigma_comparison}
\end{figure}

We compare the behaviour of $\sigma_M$ as a function of polar angle $\theta$ at $r=5,15,30\,r_g$ between models \texttt{KIa00} and \texttt{KIa09} in Figure \ref{fig:sigma_comparison}. Model \texttt{KIa00} does not have a relativistic jet and is not significantly impacted by the addition of mass in the funnel since $\sigma_M < 100$ up to $r=5\,r_g$. Model \texttt{KIa09} on the other hand is subject to density floors at  radii as large as $r<15\,r_g$. The gas density $\rho$ and magnetization $\sigma_M$ profiles are flat in $\theta$ where the ceiling is active (i.e. see the profile at $r=5\, r_g$). The radiation properties ($L_{\rm{bol}}$ and $\eta_{\rm{rad}}$) in these regions becomes dependent on the choice of ceiling model, as we have illustrated in Section \ref{sec:KIa09}.

\bsp	
\label{lastpage}
\end{document}